\documentclass[useAMS,usenatbib,usegraphicx]{mn2e}

\def\gtsim {>\kern-1.2em\lower1.1ex\hbox{$\sim$}~}   
\def\ltsim {<\kern-1.2em\lower1.1ex\hbox{$\sim$}~}   

\def \apj {ApJ}
\def \apjs {ApJS}
\def \aj  {AJ}
\def \aap {A\&A} 
\def \mnras {MNRAS}
\def \pasj {PASJ}
\def \araa {ARA\&A}

\def \cite#1{\citealt{#1}}

\title[Simulations of Cosmic Chemical Enrichment]{Simulations of Cosmic Chemical Enrichment}
\author[Chiaki Kobayashi, Volker Springel, and Simon D. M. White]{Chiaki Kobayashi$^{1}$\thanks{E-mail: chiaki@th.nao.ac.jp}, Volker Springel$^{2}$, and Simon D. M. White$^{2}$\\
  $^{1}$National Astronomical Observatory of Japan, 2-21-1 Osawa, Mitaka-shi, Tokyo 181-8588, Japan\\
  $^{2}$Max-Planck-Institute for Astrophysics, Karl-Schwarzschild-Str. 1,
  D-85741 Garching, Germany}

\begin{document}

\date{Accepted ----. Received 2006}

\pagerange{\pageref{firstpage}--\pageref{lastpage}} \pubyear{2006}

\maketitle

\label{firstpage}

\def \cite#1{\citealt{#1}}

\begin{abstract}
  Using a new numerical model for cosmic chemical evolution, we study the
  influence of hypernova feedback on the star formation and metal enrichment
  history of the universe.  For assumptions which produce plausible results in
  idealized collapse models of individual galaxies, our cosmological
  simulations of the standard $\Lambda$CDM cosmology show a peak of the cosmic
  star formation rate at $z\sim4$, with $\sim 10\%$ of the baryons turning
  into stars.  We find that the majority of stars in present-day massive
  galaxies formed in much smaller galaxies at high redshifts, giving them a
  mean stellar age as old as $10$ Gyr, despite their late assembly times. The
  hypernova feedback drives galactic outflows efficiently in low mass galaxies, and these
  winds eject heavy elements into the intergalactic medium.  The ejected
  baryon fraction is larger for less massive galaxies, correlates well with
  stellar metallicity, and amounts to $\sim 20\%$ of all baryons in total. The
  resulting enrichment history is broadly consistent with the observed abundances of Lyman break galaxies, of damped Lyman $\alpha$ systems, and of the intergalactic medium.
The
  metallicity of the cold gas in galaxies increases with galaxy mass, which is comparable to observations with a significant scatter.
  The stellar mass-metallicity relation of the observed galaxy population is
  well reproduced by the simulation model as a result of the mass-dependent
  galactic winds.  However, star formation does
  not terminate in massive galaxies at late times in our model, and too few
  dwarf galaxies are still forming stars.  These problems may be due to a lack
  of resolution, to inappropriate modelling of supernova feedback, or to a
  neglect of other feedback processes such as active galactic nuclei.
\end{abstract}


\section{INTRODUCTION}
\label{sec:intro}

While the evolution of the dark matter in the standard $\Lambda$CDM cosmology
is reasonably well understood, the evolution of the baryonic component is much
less certain because of the complexity of the relevant physical processes, such
as star formation and feedback.  One approach to study the dynamics of baryons
in galaxy formation is to use semi-analytic techniques (e.g., \cite{kau93};
\cite{col94}) that combine the growth of dark matter halos with simplified
parameterizations of the baryonic physics.  While this allows the construction
of theoretical models for galaxy formation and chemical enrichment with a
large dynamic range, the validity of the assumptions must be ultimately
justified by observations and by more detailed theoretical work. We here
adopt the other common approach to study galaxy formation, which is to use
direct hydrodynamical simulations. Complementary to the semi-analytic models,
they treat the dynamics of baryons in a much more detailed fashion.

Many simulation codes have been developed for studying galaxy formation and
evolution, not only of isolated systems (e.g., \cite{kat92}, \cite{mih94},
\cite{ste94}, \cite{nak03}, \cite{kaw03}, \cite{kob04}) but also for
cosmological simulations of individual galaxies (e.g.,
\cite{nav94}) or of the galaxy population as a whole (\cite{cen99},
\cite{spr03}). A robust result has been that with the commonly employed,
schematic star formation criteria alone, i.e., converging gas flow, local
Jeans instability, and short cooling times, the predicted star formation rates
(SFRs) are higher than what is compatible with the observed luminosity
density.  Thus feedback mechanisms are in general invoked to reheat gas and
suppress star formation. We include both supernova and hypernova feedback in
our hydrodynamical model in this paper.  Supernovae inject both thermal energy
and heavy elements into the interstellar medium, and by means of
supernovae-driven galactic winds, some of these metals can escape from
galaxies into the intergalactic medium.

There is observational evidence for galactic winds in star forming galaxies
both locally and at high redshifts (\cite{hec00}; \cite{pet01}; \cite{mar02};
\cite{maz02}; \cite{ohy02}).  The existence of heavy elements in the
intracluster medium (ICM) and intergalactic medium (IGM) also requires such an
ejection mechanism.  The mass-metallicity relation and also the
color-magnitude relation of early-type galaxies have been explained as the
the result of loss of heavy elements in galactic winds (e.g., \cite{lar74},
\cite{ari87}).  The relation between the iron mass in the ICM and the
luminosity of early-type galaxies (\cite{arn92}) may also suggest metal
ejection from early-type galaxies.

Since the presence of heavy elements can increase gas cooling substantially
(\cite{sut93}), consistent simulations of galaxy formation must include a
treatment of chemical enrichment as well as energy feedback.  However,
especially with respect to chemical enrichment, most existing hydrodynamical
models are too simplistic to be compared in detail with the real Universe.
There are only a few studies that simulate the detailed chemical enrichment by
both Type II and Ia supernovae (SNe II and SNe Ia) (\cite{rai96},
\cite{car98}, \cite{mos01}, \cite{nak03}, \cite{kaw03}, \cite{tor04},
\cite{kob04}).
Different types of supernovae produce different heavy elements on different timescales. Although
the most abundant heavy element is oxygen which accounts for half of the solar
metallicity, this element is hardly detected in observations of stars.  Iron,
which is another fundamental element observed in the damped Lyman $\alpha$
(DLA) systems and the ICM, is mainly produced by SNe Ia.  Recently, it has
been confirmed that supernovae with ten times the standard kinetic energy
exist and produce a certain amount of iron, not only by the observation of
individual supernovae (e.g., \cite{nom02}) but also by the elemental abundance
ratios of Milky Way stars (\cite{kobnom06}).  Even with the Salpeter initial
mass function (IMF), three times more energy is ejected on average when such
hypernovae (HNe) are included, which therefore increase the overall strength
of supernova feedback substantially.

In the context of hierarchal galaxy formation, a number of important questions
should be explained by simulation models, including (1) the existence of
massive galaxies at high redshifts (e.g., \cite{fos04}), (2) the old stellar
populations of massive elliptical galaxies (e.g., \cite{bow92}), (3) the
``down-sizing'' effect of star formation (e.g., \cite{cow96}, \cite{kod04}),
and (4) the number of spiral galaxies and the detailed morphological mix of
galaxies.  With our simulations, we can only address questions (2) and (3)
because of limitations in volume and resolution.  Note that while more massive
halos tend to form at later times in hierarchal clustering, this does not
necessarily imply that massive galaxies have younger stars than low-mass
galaxies.  This is because the present-day massive galaxies may have formed by
the merging of smaller galaxies, i.e.~the assembly time of these systems needs
not to be the same as the formation time of their stars. To study this
question, we will directly analyze when and where stars form in our
simulations and we discuss the ages of stellar populations of massive galaxies
formed in our models.

It is not clear that the IMF is universal, and
simulations of the formation of Population III stars appear to suggest a
flatter IMF or a lack of low-mass stars in low-metallicity environments (e.g.,
\cite{bro04}).  This uncertainty means that the SFR and the stellar mass
cannot be uniquely obtained from the observed luminosity.  However, since a
flatter IMF produces more heavy elements, the metal enrichment history of
galaxies can put constraints on their star formation history and IMF. In fact,
the observed weak chemical evolution of the IGM (\cite{son01}, \cite{sch03})
and of DLA systems (\cite{pet97}, \cite{pro03}) may conflict with a flatter
IMF. A detailed treatment of chemical
enrichment is required in cosmological simulations.  To this end we will
analyze where and how heavy elements are produced and distributed in the gas
phase and in stars, using a Salpeter IMF for definiteness.

In this paper, we simulate the evolution of gas and stellar systems from
cosmological initial conditions, following chemical enrichment from SNe II,
SNe Ia, and HNe.  We use the {\small GADGET-2} code by \citet{spr05}, in
which we introduce metal-dependent cooling rates (\cite{sut93}) and the
chemical enrichment model of Kobayashi (2004, hereafter K04).  The feedback
scheme is slightly modified from K04, as described in Section~2.  In
Section~3, as a test of our code, we simulate galactic winds from dwarf spiral
galaxies and examine their parameter dependencies.  Our cosmological
simulations are described in Section~4, where we also discuss their
predictions for the star formation history, galactic winds, and the chemical
enrichment.  Section~5 gives our conclusions.

\begin{figure*}
\begin{center}
\includegraphics[width=6.5cm,angle=-90]{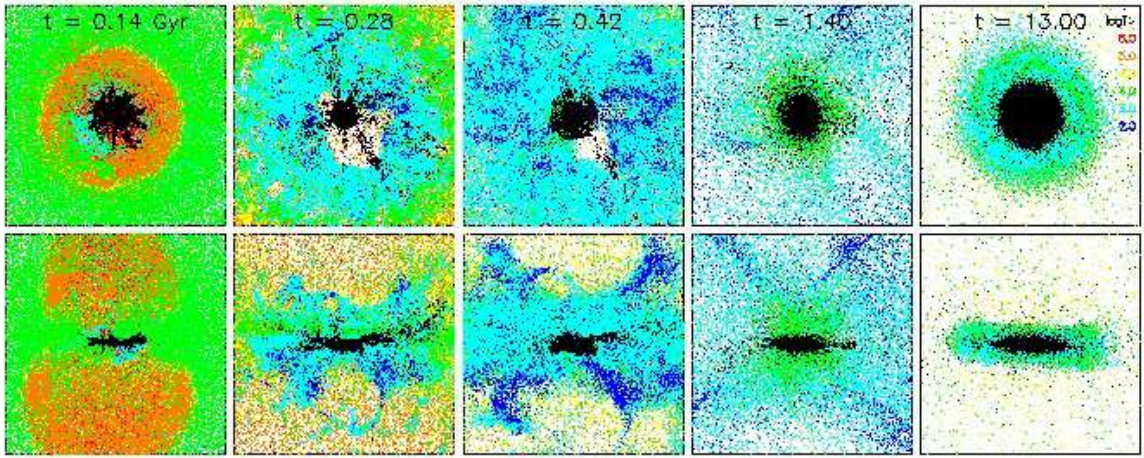}
\caption{\label{fig:diskmap}
  The time evolution of the formation of an isolated disk galaxy in a halo of
  mass of $10^{10}\, h^{-1} {\rm M}_\odot$. The black points show star particles, while the gas
  particles are colour-coded according to their temperature. Each panel is
  $20$ kpc on a side. The upper row shows face-on projections, the lower row
  gives edge-on views.  }
\end{center}
\end{figure*}

\section{Simulation Model}

To construct a self-consistent three-dimensional chemodynamical model, we have
introduced various physical processes associated with the formation and
evolution of stellar systems into the parallel tree-SPH code {\small GADGET-2}
(\cite{spr05}, \cite{spr00}), which is fully adaptive with individual smoothing
lengths and timesteps and uses an entropy-conserving formulation of SPH
(\cite{spr02}).  We include radiative cooling, and photo-heating by a uniform
and evolving UV background radiation (\cite{haa96}). Star formation, feedback
from Type II and Ia supernovae (SNe II and SNe Ia), stellar winds (SWs), and
chemical enrichment are modelled as well, with an implementation that is close
to the one in K04. In brief, the characteristics of this part of our code may
be summarized as follows.

\begin{enumerate}
\item {\bf Radiative cooling} is computed using a metallicity-dependent
  cooling function.  For primordial gas ([Fe/H] $<-5$), we compute the cooling
  rates using the two-body processes of H and He, and free-free emission, as
  in \citet{kat96}.  For metal enriched gas ([Fe/H] $\ge -5$), we use a
  metallicity-dependent cooling function computed with the MAPPINGS III
  software (\cite{sut93}).  In this cooling function, the elemental abundance
  ratios are set to be constant for given [Fe/H] according to the relations
  found in the solar neighbourhood. [O/Fe]$=0.5$ for Galactic halo stars for
  [Fe/H] $\le -1$, and solar values for [Fe/H] $\ge 0$. We interpolate between
  these values for $-1<$ [Fe/H] $<0$.

\item Our {\bf star formation} criteria are the same as in \citet{kat92}: (1)
  converging flow, $(\nabla \cdot \mbox{\boldmath$v$})_i < 0$; (2) rapid
  cooling, $t_{\rm cool} < t_{\rm dyn}$; and (3) Jeans unstable gas, $t_{\rm
    dyn} < t_{\rm sound}$.  The star formation timescale is taken to be
  proportional to the dynamical timescale ($t_{\rm sf} \equiv
  \frac{1}{c_*}t_{\rm dyn}$), where $c_*$ is a star formation timescale
  parameter which we set to $0.1$ (\cite{kob05}).  For every timestep $\Delta
  t$, we draw a random number $p$ between 0 and 1, and provided it fulfils $p
  \le \frac{m}{m_{\rm g,0}/N_*} \left(1 - \exp\left[-\frac{\Delta t}{t_{\rm
          sf}}\right] \right)$, a fractional part of the mass of the gas
  particle turns into a new star particle.  The mass of the new star particle
  is given as $m_{*,0}=m_{\rm g,0}/N_*$ with $N_*=2$.  Note that an individual
  star particle has a typical mass of $\sim 10^{7} M_\odot$, i.e.~it does not
  represent a single star but an association of many stars.  The masses of
  the stars associated with each star particle are distributed according to an
  initial mass function (IMF).  We adopt a power-law IMF, $\phi(m) \propto
  m^{-x}$ (the slope $x=1.35$ gives the Salpeter IMF), which is assumed to be
  independent of time and metallicity.  We limit the IMF to the mass range
  $0.1M_\odot \le m \le 120M_\odot$.

\item We do not adopt the instantaneous recycling approximation when
  accounting for {\bf feedback}.  Instead, we treat star particles as evolving
  stellar populations which eject thermal energy $E_e$, gas mass $E_m$, and
  heavy elements $E_{z_i}$ from SWs, SNe II, and SNe Ia, as a function of
  time.  The release of energy and heavy elements is distributed to a constant
  number $N_{\rm FB}$ of surrounding gas particles.  We will show results for
  two cases: In our `SN feedback' model, the SN II energy and yields are the
  same as adopted in K04.  In the `HN feedback' model, we assume instead a constant
  hypernova fraction $\epsilon_{\rm HN}=0.5$ for progenitor masses $M
  \ge 20\,{\rm M}_\odot$, and adopt the mass-energy relation; $30$, $20$, $10$, $10 \times
  10^{51}$ erg for $40$, $30$, $25$, $20\,{\rm M}_\odot$ (\cite{kobnom06}).  The ejected
  energy of each SW, SN II, and SN Ia are $\sim 0.2 \times 10^{51}$ erg
  depending on metallicity, $1 \times 10^{51}$ erg, and $1.3 \times 10^{51}$
  erg, respectively.

We distribute this feedback energy in purely thermal form, although a fraction
of it could, in principle, be distributed in kinetic form as a velocity
perturbation to the gas particles (see \cite{nav93}).  The energy increase of
a gas particle $i$ due to a star particle $j$ is calculated in each timestep
$\Delta t_j$ as
\begin{equation}
\frac{{\rm d}u_i}{{\rm d}t}=\int_{t-\Delta t_j}^{t} {\rm d}t\;E_{e,j}(t,Z_j) \, W(r_{ij}) / \sum_{k=1}^{N_{\rm FB}} W(r_{jk}).
\end{equation}
The feedback neighbour search needs to be done twice in order to ensure proper
mass and energy conservation; first to compute the sum of weights for the
normalization, and a second time for the actual distribution.  Roughly
speaking, since we assume constant $N_{\rm FB}$, the energy increase per mass
is almost constant and independent of resolution; $E_e/m_{\rm g}/N_{\rm FB}
\sim m_{*,0} {\cal{R}}_{\rm SN} 10^{51} {\rm erg} \, /m_{\rm g}/N_{\rm FB}
\propto 1/N_*/N_{\rm FB}$.  (Naively, the mass of gas particles are decreasing
because of star formation.) The feedback radius that was assumed to be
constant in the {\small GRAPE-SPH} code of K04 is now variable depending on
the local gas density.  With higher resolution, constant $N_{\rm FB}$ gives
smaller feedback radius.

For the metals, the mass- and metallicity-dependent nucleosynthesis yields of
SNe II and SNe Ia are taken from Kobayashi et al. (2006) and Nomoto et al.
(1997b), respectively.  The progenitor mass ranges of SWs and SNe II are
$8-120\,{\rm M}_\odot$ and $8-50\,{\rm M}_\odot$, respectively.  For SNe Ia,
we adopt the single degenerate scenario with the metallicity effect
(\cite{kob98}, 2000), where the progenitors are Chandrasekhar white dwarfs
(WDs) with an initial mass of $3-8\,{\rm M}_\odot$, and the lifetimes are
determined from the lifetimes of the secondary stars with $0.9-1.5\,{\rm
  M}_\odot$ and $1.8-2.6\,{\rm M}_\odot$ for the red-giant (RG) and
main-sequence (MS) systems, respectively.  The fractions of WDs that
eventually produce SNe Ia are adjusted to match the chemical evolution
constraints of the Milky Way Galaxy as [$b_{\rm RG}=0.02, b_{\rm MS}=0.04$].

\item The {\bf photometric evolution} of a star particle is identical to the
  evolution of a simple stellar population.  Spectra $f_\lambda$ are taken
  from \citet{kod97} as a function of age $t$ and metallicity $Z$.

\end{enumerate}

\begin{table*}
\begin{tabular}{ll|cccc|ccccc}
\hline
\footnotesize
Feedback & Z-cooling & $M_{\rm vir}$ & $N_{\rm gas}$ & $\epsilon_{\rm gas}$ & $N_{\rm FB}$ & $f_{*}$ & $f_{\rm w}$ & $f_{\rm Z,w}$ & $\log Z_{*}/Z_\odot$ & $\log Z_{\rm w}/Z_\odot$\\ 
\hline
n     & n & $1$ & $40000$ & $0.05$ & -    & 0.27 &  &  &  & \\
SN    & n & $1$ & $40000$ & $0.05$ & $72$ & 0.08 & 0.68 & 0.44 & -0.42 & -1.28 \\
SN    & y & $1$ & $40000$ & $0.05$ & $72$ & 0.27 & 0.54 & 0.03 & -0.09 & -1.82 \\
SN+HN & n & $1$ & $40000$ & $0.05$ & $72$ & 0.02 & 0.58 & 0.60 & -1.52 & -1.70 \\
SN+HN & y & $1$ & $40000$ & $0.05$ & $72$ & 0.17 & 0.49 & 0.25 & -0.43 & -1.12 \\
\hline
SN    & y & $1$ & $40000$ & $0.05$ & $36$ & 0.08 & 0.80 & 0.21 & -0.23 & -1.67 \\
SN    & y & $1$ & $40000$ & $0.05$ & $144$ & 0.10 & 0.73 & 0.31 & -0.31 & -1.36 \\
SN    & y & $1$ & $10000$ & $0.08$ & $72$ & 0.19 & 0.67 & 0.17 & -0.13 & -1.30 \\
SN    & y & $1$ & $160000$ & $0.0315$ & $72$ & 0.25 & 0.47 & 0.026 & -0.10 & -1.84 \\
\hline
n     & n & $100$ & $40000$ & $0.25$ & -    & 0.59 &  &  &  & \\
SN    & n & $100$ & $40000$ & $0.25$ & $72$ & 0.47 & 0.028 & 0.0003 & -0.08 & -2.22 \\
SN    & y & $100$ & $40000$ & $0.25$ & $72$ & 0.52 & 0.011 & 0 & -0.02 & - \\
\hline
\end{tabular}
\caption{Properties of different simulations of
isolated disk galaxy formation, at time $t=13$ Gyr. In all runs, we adopted 
$c=0.1$. The total mass and gravitational softening length are given 
in units [$h^{-1}10^{10}{\rm M}_\odot$] and [$h^{-1}$ kpc], respectively.
The `wind gas' is here defined as the gas particles that are outside $r>2r_{200}$.
$f_{\rm w}$ and $f_{\rm Z,w}$ denote the wind fraction relative to the total baryon mass and the ejected metal fraction relative to the total amount of metals produced, respectively.
}
\label{tab:disk}
\end{table*}

\section{ISOLATED DISKS}
\label{sec:disk}

As a test of our code, we show the evolution of an isolated disk galaxy in a
static dark matter potential.  We construct an NFW-halo (\cite{nav96}) with
$10\%$ gas content in virial equilibrium for total masses of $M_{\rm
  vir}=10^{10}\, h^{-1} {\rm M}_\odot$ and $10^{12}\, h^{-1} {\rm M}_\odot$.
The initial angular momentum corresponds to a spin parameter $\lambda=0.1$ and
is distributed within the halo under the assumption that the specific angular
momenta of spherical shells are all aligned (\cite{bul01}).  For the
$10^{10}\, h^{-1} {\rm M}_\odot$ model, we work with a numerical resolution of
$N_{\rm gas}=10000$, $40000$, and $160000$ particles, corresponding to a gas
particle mass of $1\times10^5$, $2.5\times10^4$, and $6.25\times10^3\, h^{-1}
{\rm M}_\odot$, respectively.  The adopted gravitational softening length is
given in Table~\ref{tab:disk}.

\begin{figure}
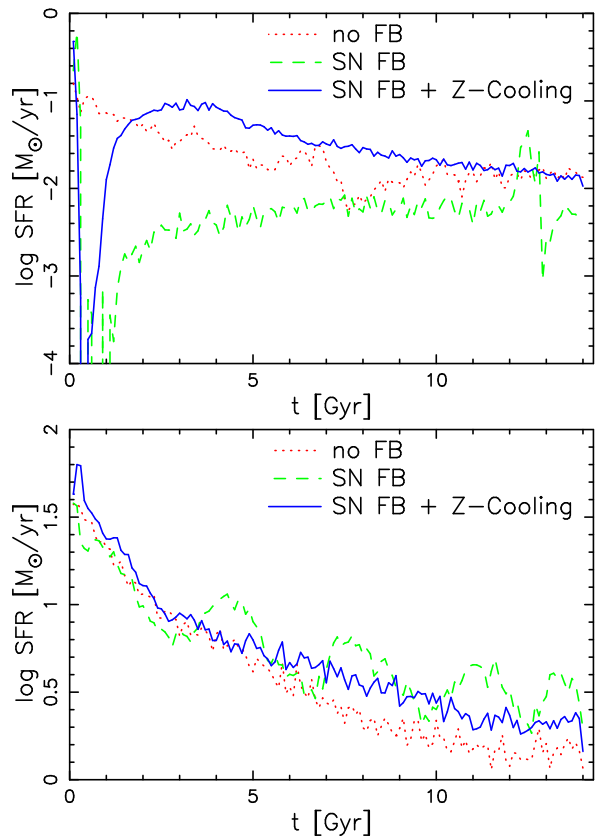

\begin{center}
\includegraphics[width=5.5cm,angle=-90]{fig2a.ps}
\includegraphics[width=5.5cm,angle=-90]{fig2b.ps}
\caption{\label{fig:disk_sfr}
  Star formation rates in isolated halos with total mass $10^{10}\,h^{-1}\,{\rm
    M}_\odot$ (upper panel) and $10^{12}\,h^{-1}\,{\rm M}_\odot$ (lower panel).  The
  dotted, dashed, and solid lines show the results with no feedback, feedback,
  and feedback plus metal-dependent cooling.}
\end{center}
\end{figure}

\begin{figure}
\begin{center}
\includegraphics[width=7.8cm]{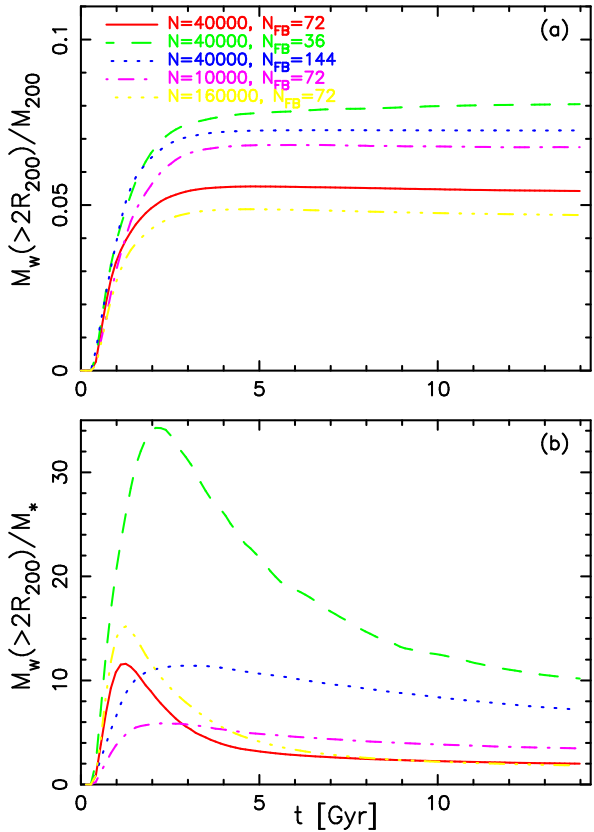}
\caption{\label{fig:disk_win1}
  The time evolution of the wind mass in $10^{10}\,h^{-1} {\rm M}_\odot$ disks
  with different resolution and feedback parameters for the model with
  feedback and metal-dependent cooling; $N=40000, N_{\rm FB}=72$ (solid line),
  $N_{\rm FB}=36$ (dashed line), $144$ (dotted line), $N=10000$ (dot-dashed
  line), and $N=160000$ (dash-dot-dot-dot line).  In the lower panel, the wind
  mass is normalized by the stellar mass at each time, which corresponds to
  the wind efficiency.  }
\end{center}
\end{figure}

Figure~\ref{fig:diskmap} shows the time evolution of gas and star particles in
the $10^{10}\,h^{-1}{\rm M}_\odot$ halo for the case with SN feedback
($N=160000$ resolution).  After the start of the simulation, the gas is
allowed to cool radiatively, making the gas in the centre of the halo quickly
lose its pressure support such that it settles into a rotationally supported
disk that grows from inside out.  The rapid initial growth of the gas disk 
induces an initial starburst.  Figure~\ref{fig:disk_sfr} compares the
star formation rates (SFRs) for the models without feedback (dotted line),
with feedback (dashed line), and with metal-dependent cooling (solid line).
The SFR is peaked at $t\sim0.07$ Gyr with maxima of 0.25, 0.45, and 0.8 ${\rm
  M}_\odot$ yr$^{-1}$, respectively.  Since the shortest lifetime of SNe II is
$\sim 4 \times 10^6$ yr, star formation is affected by the feedback during
this phase.  Although the adopted star formation scheme is different, our
model without feedback is similar to that of \citet{spr03}.  With
metal-dependent cooling, star formation is more efficient with a twice larger
peak SFR.

As soon as stars form in the centre, thermal energy of SNe II is injected into
the inter-stellar medium.  The energy from a star particle is isotropically
distributed to the surrounding gas particles in our model.  Because of the
dense disk formed and the non-isotropic infall of gas, the low-density hot-gas
region expands in a bipolar flow ($t\sim0.14\,{\rm Gyr}$) as result of the
energy input.  In the disk plane, the hot gas region expands and forms a dense
shell where stars keep on forming.  The energy from these stars can quickly
propagate to the surrounding low-density region.  After the galactic wind
forms ($t\sim0.28\,{\rm Gyr}$), the gas density becomes so low at the centre
that star formation is terminated for a while.  Because of radiative cooling,
a part of the ejected gas however returns, again settling in the disk where it
fuels new star formation.  This secondary star formation is not as strong as
the initial starburst, and takes place in a more continuous fashion.  Although
some small bubbles are forming in the galaxy in this stage, not much gas is
ejected by them from the disk.

Supernovae eject not only thermal energy but also heavy elements into the
interstellar medium.  Since the cooling function depends strongly on
metallicity, the star formation history is altered when metal-dependent
cooling is accounted for.  This is important especially for the secondary star
formation when the gas has become enriched.  Indeed, when metal-dependent
cooling is included, the secondary star formation is much stronger than
without it.  Compared with the no feedback case, the final stellar mass
decreases by a factor of $3$ if we include the feedback, but increases back to
the original value if we add metal-dependent cooling as well.  In
Table~\ref{tab:disk}, we summarize the properties of the galaxies in the
different cases after a time of 13~Gyr.

For the massive halo with $10^{12}\,h^{-1} {\rm M}_\odot$ (the lower panel of
Fig.~\ref{fig:disk_sfr}), a galactic wind never occurs, and star formation
takes place continuously.  When we include feedback, the SFR shows repeated
bursting behaviour.  However, with metal-dependent cooling, star formation
again takes place smoothly.  This is because the enhancement of the star
formation due to the existence of heavy elements is larger than the
suppression by thermal feedback.

We note that these initial conditions are somewhat artificial since
galaxies are not born as pure gas disks in reality. We adopt these
initial conditions in order to explore the effects of feedback in a
clear way, without having to deal with the complications of the
hierarchical formation of galaxies. In reality, massive galaxies form
through the merging of subgalaxies, and galactic winds can occur from
the subgalaxies before merging.  Weather the ejected materials fall onto
the final galaxies or not is another issue, and should depend on the
details of the merging histories.  In Section 4, we will adopt more
realistic initial conditions appropriate for cosmological simulations
where the hierarchical formation is fully taken into account.

\begin{figure}
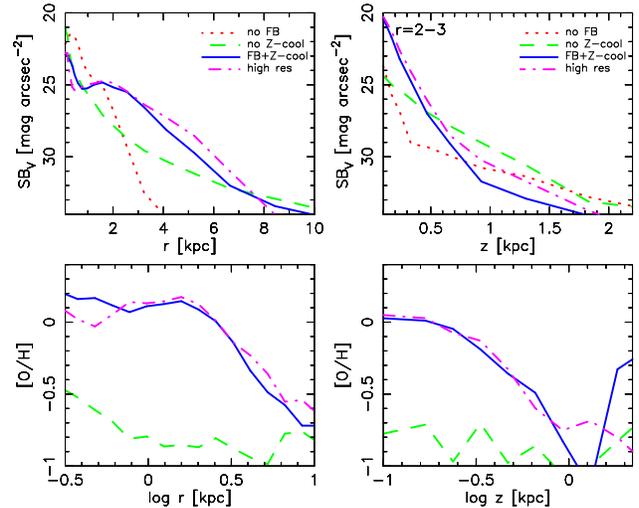

\begin{center}
  \includegraphics[height=6.7cm]{fig4a.ps}
  \includegraphics[height=6.7cm]{fig4b.ps}
\caption{\label{fig:disk_cgrad}
  Surface brightness profiles as a function of projected radius (left) and
  height (right).  The lower panels show the oxygen abundance gradients.  }
\end{center}
\end{figure}

\begin{figure*}
\begin{center}
  \includegraphics[width=13cm,angle=-90]{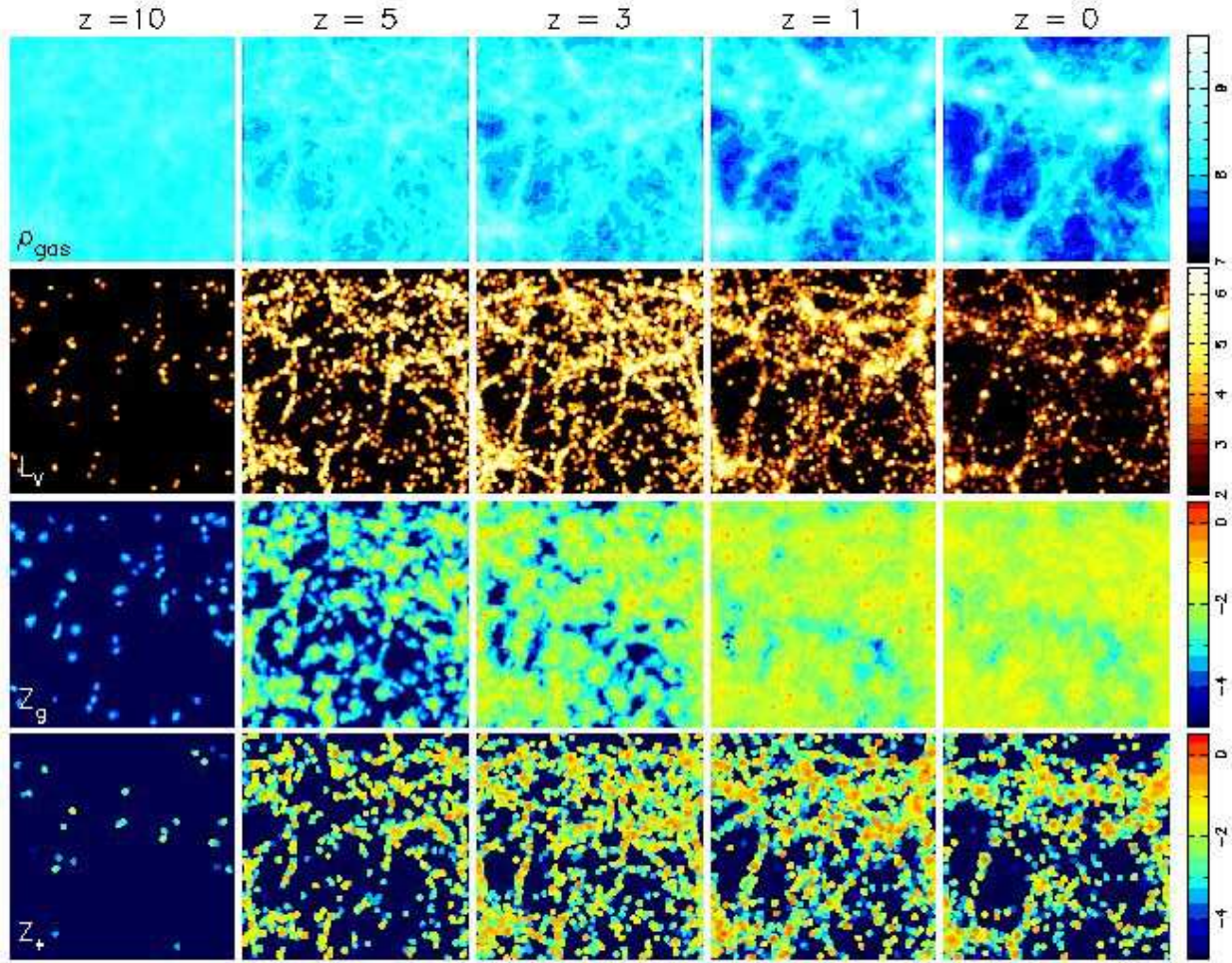}
  \vspace*{0.1cm}
\caption{\label{fig:cosmicmap}
  The time evolution of our cosmological simulation in a periodic box $10
  \,h^{-1}\,{\rm Mpc}$ on a side. We show the projected density of gas (first
  row), stellar V-luminosity (second row), gas metallicity $\log Z_{\rm g}/Z_\odot$ (third row) and stellar
  metallicity $\log Z_*/Z_\odot$ (forth row) for the high
  resolution simulation ($N=96^3, N_{\rm FB}=405$). }
\end{center}
\end{figure*}

The strength of the galactic winds depends on 
i) the feedback parameter and ii) the numerical
resolution.  Figure~\ref{fig:disk_win1} shows the time evolution of the wind
mass normalized by the total mass (upper panel) and the stellar mass (lower
panel) as a function of time.  The wind particles are here defined as those
gas particles that are outside $r>2\,r_{200}$.  In our standard model, almost
half of the gas is ejected by the wind ($f_{\rm w}=0.54$).  In our feedback
scheme, the feedback neighbour $N_{\rm FB}$ controls the ejected energy from
one star particle to one gas particle.  Once the gas particles are heated to
$T \gtsim 3 \times 10^5$ K where the cooling rate becomes low, they
can avoid rapid cooling.  Therefore, the feedback effect becomes stronger with
smaller $N_{\rm FB}$.  With $N_{\rm FB}=36$, the wind mass increases to give
$f_{\rm w}=0.8$, and the wind efficiency normalized by the stellar mass is
$\sim 5$ times larger than for the $N_{\rm FB}=72$ case.  On the other hand,
with $N_{\rm FB}=144$, the distributed energy is too low to generate a
galactic wind.  In addition, a large value of $N_{\rm FB}$ corresponds to a
very large feedback radius ($\sim 10$ kpc in these simulations), which causes
unrealistically large mixing of gas.

A comparable amount of heavy elements can be ejected by the wind with the
feedback. However, with the metal-dependent cooling, the ejected metal
fraction relative to the total amount of metals produced is as small as a few
percent for the SN feedback case.  For the case with HNe, the ejected metal
fraction is as large as $f_{\rm Z,w}=0.25$.  This is because more metals are ejected with
more energy, and enriched gas can be heated enough to prevent rapid cooling.
In this case, the wind metallicity is $\log Z/Z_\odot \sim -1$ while the
stellar metallicity is $\sim -0.4$ (see Table~\ref{tab:disk}).

In principle, our feedback scheme does not depend on numerical resolution
because the ejected energy per unit mass from star particles is always the
same.  However, the star formation criteria may introduce some resolution
dependence.  In our disk formation simulations, the feedback efficiency depends
on the strength of the initial star burst.  We find that the SFR during the
initial starburst is smaller with $N=160000$ than $N=40000$.  This is because
the Jeans instability criterion is not as easily satisfied in high resolution
simulations with smaller smoothing length.  $N=10000$ is too small to simulate
the galactic wind properly.  As a result, the wind mass decreases with higher
resolution. However, when we normalized the wind mass by the stellar mass, 
the present wind efficiency is the same.

The feedback also affects the structure of the stellar disk.  In
Figure~\ref{fig:disk_cgrad}, we show the surface brightness profiles of the
stars in the $10^{10}\,h^{-1}{\rm M}_\odot$ halo as functions of the projected
radius (left panels) and the height at $r=2-3$ kpc (right panels).  Without
feedback (dotted line), the disk is small and thin with scale lengths of
$r_{\rm e}=0.5$ and $h=0.1\,{\rm kpc}$, respectively.  With feedback and
metal-dependent cooling (solid line), the disk size becomes larger ($r_{\rm
  e}=1.3$), while the thickness does not change ($h=0.13$).  At late times,
star formation takes place mainly in the outer regions ($r\sim2-3$ kpc) in
this case.  Thus the surface brightness profile shows a bump around $r\sim2$
kpc.  Strong metallicity gradients are seen at $r>2\,{\rm kpc}$ and
$z>0.2\,{\rm kpc}$.  Without metal-dependent cooling (long-dashed line), the
disk size is smaller and the thickness is larger.  In this case, star
formation takes place in the centre at later times, and no metallicity
gradient is formed.  Numerical resolution does not affect these results
(dot-dashed line).

\section{Cosmological simulations}

We focus on a $\lambda$CDM cosmological model with parameters $H_0=70\,{\rm km
  \,s^{-1} Mpc^{-1}}$, $\Omega_m=0.3$, $\Omega_\Lambda=0.7$, $\Omega_{\rm
  b}=0.04$, $n=1$, and $\sigma_8=0.9$.  The initial conditions are set up in a
$10\, h^{-1}{\rm Mpc}$ cubic box with periodic boundary conditions, employing
an equal number of dark matter and gas particles.  We work with two different
resolutions; $N_{\rm DM}=N_{\rm gas}=54^3$ and $96^3$.  The mass of a dark
matter particle is $4.58 \times 10^8\,{\rm M}_\odot$ and $8.16 \times
10^7\,{\rm M}_\odot$ and the mass of a gas particle is $7.05 \times 10^7 {\rm
  M}_\odot$ and $1.25 \times 10^7\,{\rm M}_\odot$ for these two resolution,
respectively.  The adopted gravitational softening lengths are given in
Table~\ref{tab:cosmic}. For gas particles we adopt half the value adopted for
dark matter.  With several low-resolution simulations we examine dependencies
on our two most important model parameters, the feedback neighbour number
$N_{\rm FB}$ and the star formation timescale $c_*$.

In order to identify galaxies, we apply the same method as \citet{spr03b}.  A
friend-of-friends (FOF) group-finding algorithm is applied to the dark matter
particles, using a fixed comoving linking length equal to $0.2$ times the mean
inter-particle spacing of the dark matter particles.  Then gas and star
particles are associated with the nearest dark matter particles.  We discard
all groups with fewer than $32$ dark matter particles.  The intergalactic
medium is coarsely defined as the particles that do not belong to any
identified halo. We note that our results are insensitive to the detailed
parameters of the FOF group finding methods.

Because of practical limitations with respect to the available
computational time, the box size of our simulations is quite
limited. Thus very massive galaxies and galaxy clusters forming at low
redshift are not included and the large-scale structure on the scale of
the box-size is affected by non-linear evolution below $z\sim2-3$.  As a
result, the simulated cosmic SFR in Figure~\ref{fig:sfr} at $z<2-3$ is probably an
underestimate of the cosmic mean expected for a larger simulation
volume.  However, the evolution of individual galaxies is expected to be
less affected by the finite box size. In particular, the
mass-metallicity relation (Figs. \ref{fig:feh_mass} and 
\ref{fig:fehs_mass}) and the wind-mass relation
(Fig. \ref{fig:wind}) should be quite robust for low and intermediate-mass galaxies,
even though very massive galaxies are not included.

Figure~\ref{fig:cosmicmap} shows the time evolution of the densities of gas,
stellar V-band luminosity, gas metallicity, and stellar metallicity for the
high resolution simulation ($N=96^3, N_{\rm FB}=405$) in our $10 h^{-1}{\rm
  Mpc}$ box.  Star formation takes place in a distributed fashion at high
redshifts, and becomes most active around $z=3$.  The distribution of stars
appears smooth at high redshifts, but concentrated at lower redshifts.
Because of the feedback, massive galaxies are surrounded by hot gas, and heavy
elements are also distributed in the intergalactic medium.  Both in the gas
phase and stars, metallicity gradients are generated in high density regions.

\begin{figure}
\begin{center}
\includegraphics[width=5.5cm,angle=-90]{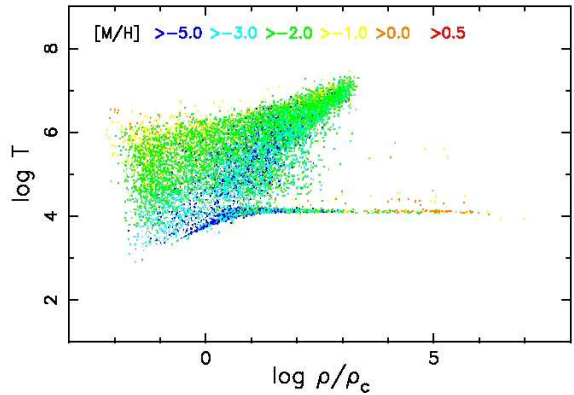}
\caption{\label{fig:rhot}
  The density-temperature diagram at $z=0$, colour-coded by the metallicity of
  gas particles. }
\end{center}
\end{figure}

\begin{figure}
\begin{center}
\includegraphics[width=5.5cm,angle=-90]{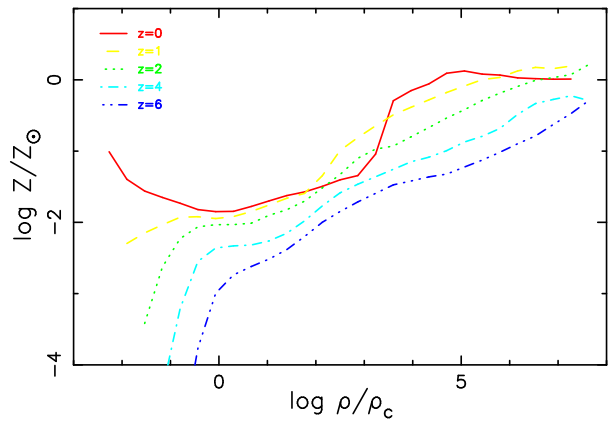}
\caption{\label{fig:rhoz}
  The metallicity-density relation at $z=0$ (solid line), $z=1$ (dashed line),
  $z=2$ (dotted line), $z=4$ (dot-dashed line), and $z=5$ (dot-dot-dot-dashed
  line).  }
\end{center}
\end{figure}

Figure~\ref{fig:rhot} shows the density-temperature diagram at $z=0$,
representing metallicities with colours.  Metal-rich gas particles populate
mainly the cold dense region, with a few particles found in the hot,
low-density region.  We note that one-tenth of particles are randomly chosen
and plotted in this diagram, and the statistics of the low-density cold gas is
much better.  Figure~\ref{fig:rhoz} shows that the relation that
the gas metallicity is higher for high density is present at all redshifts.
This is also responsible for generating the radial metallicity gradients in
galaxies both for gas and for stars.

\begin{table*}
\begin{tabular}{ll|ccccc|ccccccc}
\hline
\footnotesize
Feedback & Z-cooling & Size & $N/2$ & $\epsilon_{\rm DM}$ & $N_{\rm FB}$ & $c_*$ & $f_{\rm *}$ & $Z_{\rm g}/Z_\odot$ & $Z_*/Z_\odot$ & [Fe/H]$_*$ & [O/H]$_*$ & $f_{\rm w}$ & $f_{\rm Z,w}$ \\ 
\hline
n     & n & 10 & $56^3$ & $8.0$ & -    & $0.1$ & 0.242 & -     & -    & - & - & & \\
SN    & n & 10 & $56^3$ & $8.0$ & $72$ & $0.1$ & 0.109 & 0.034 & 0.79 & -0.02 &  0.01 & 0.11 & 0.07\\
SN    & y & 10 & $56^3$ & $8.0$ & $72$ & $0.1$ & 0.158 & 0.026 & 0.94 &  0.09 &  0.08 & 0.12 & 0.03 \\
SN+HN & n & 10 & $56^3$ & $8.0$ & $72$ & $0.1$ & 0.079 & 0.019 & 0.66 & -0.14 & -0.14 & 0.12 & 0.04 \\
SN+HN & y & 10 & $56^3$ & $8.0$ & $72$ & $0.1$ & 0.093 & 0.013 & 0.75 & -0.06 & -0.08 & 0.14 & 0.05 \\
\hline
SN+HN & y & 10 & $56^3$ & $8.0$ & $36$ & $0.1$ & 0.100 & 0.015 & 0.75 & -0.06 & -0.08 & &\\
SN+HN & y & 10 & $56^3$ & $8.0$ & $72$ & $0.02$ & 0.152 & 0.027 & 0.74 & -0.09 & -0.09 & &\\
SN+HN & y & 10 & $96^3$ & $4.5$ & $72$ & $0.1$ & 0.138 & 0.014 & 0.79 & -0.03 & -0.06 & 0.15 & 0.03 \\
SN+HN & y & 10 & $96^3$ & $4.5$ & $405$ & $0.1$ & 0.078 & 0.021 & 0.62 & -0.15 & -0.16 & 0.18 & 0.14 \\
SN+HN & y & 10 & $128^3$ & $3.4$ & $959$ & $0.1$ & 0.091 & 0.029 & 0.56 & -0.20 & -0.21 & &\\
SN+HN & y & 20 & $108^3$ & $4.5$ & $72$ & $0.1$ & 0.102 & 0.012 & 0.75 & -0.07 & -0.09 & &\\
\hline
\end{tabular}
\caption{Mean values at present epoch $z=0$.
The box size and the softening length $\epsilon$ are in [$h^{-1}$ Mpc] and [$h^{-1}$ kpc], respectively.
Wind mass is calculated from tracing the orbit of gas particles.
$f_{\rm w}$ and $f_{\rm Z,w}$ denote the wind fraction relative to the total baryon mass and the ejected metal fraction relative to the total amount of metals produced, respectively.}
\label{tab:cosmic}
\end{table*}

\subsection{Model Comparison}

\begin{figure}
\begin{center}
\includegraphics[width=8cm]{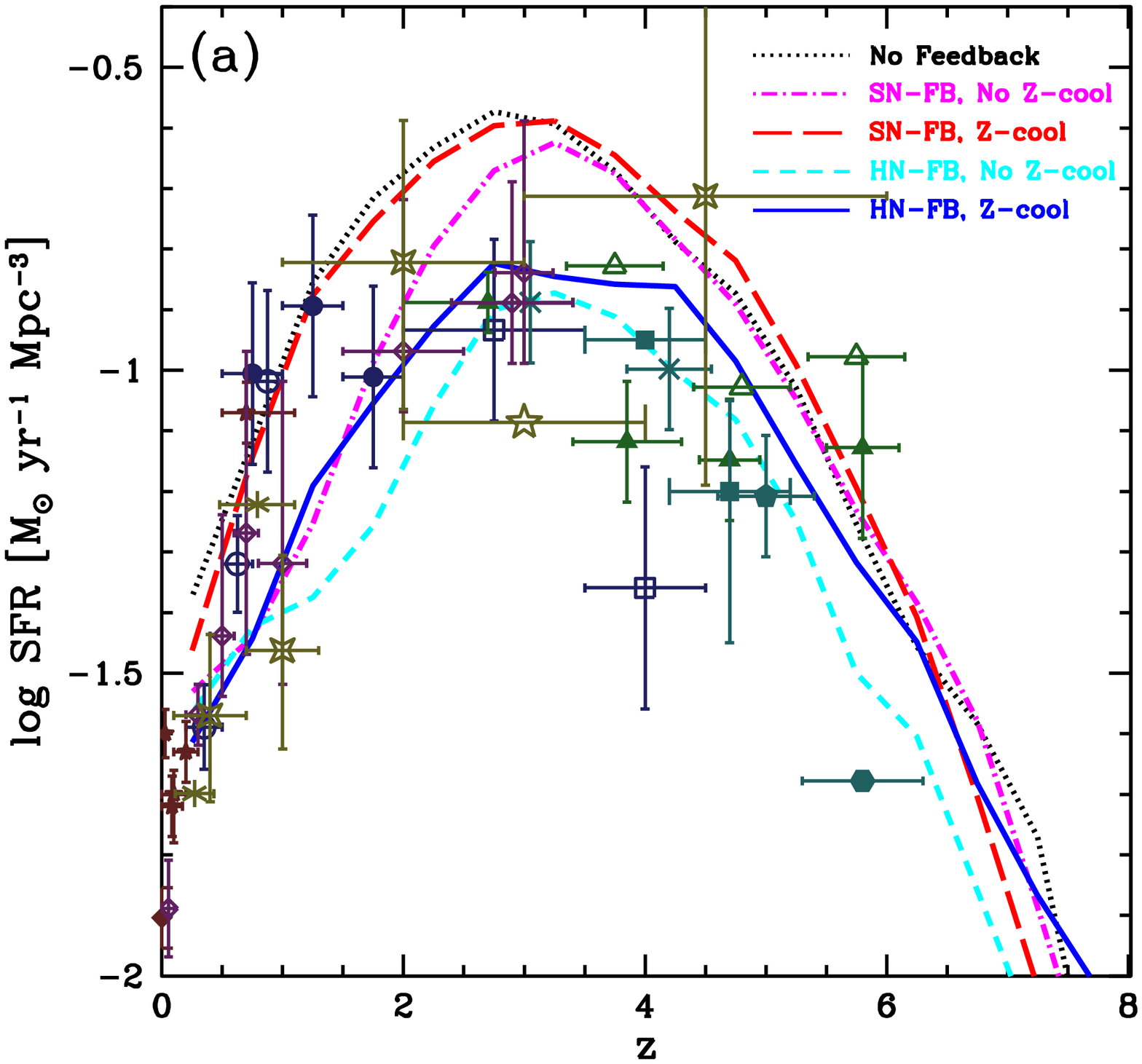}
\includegraphics[width=8cm]{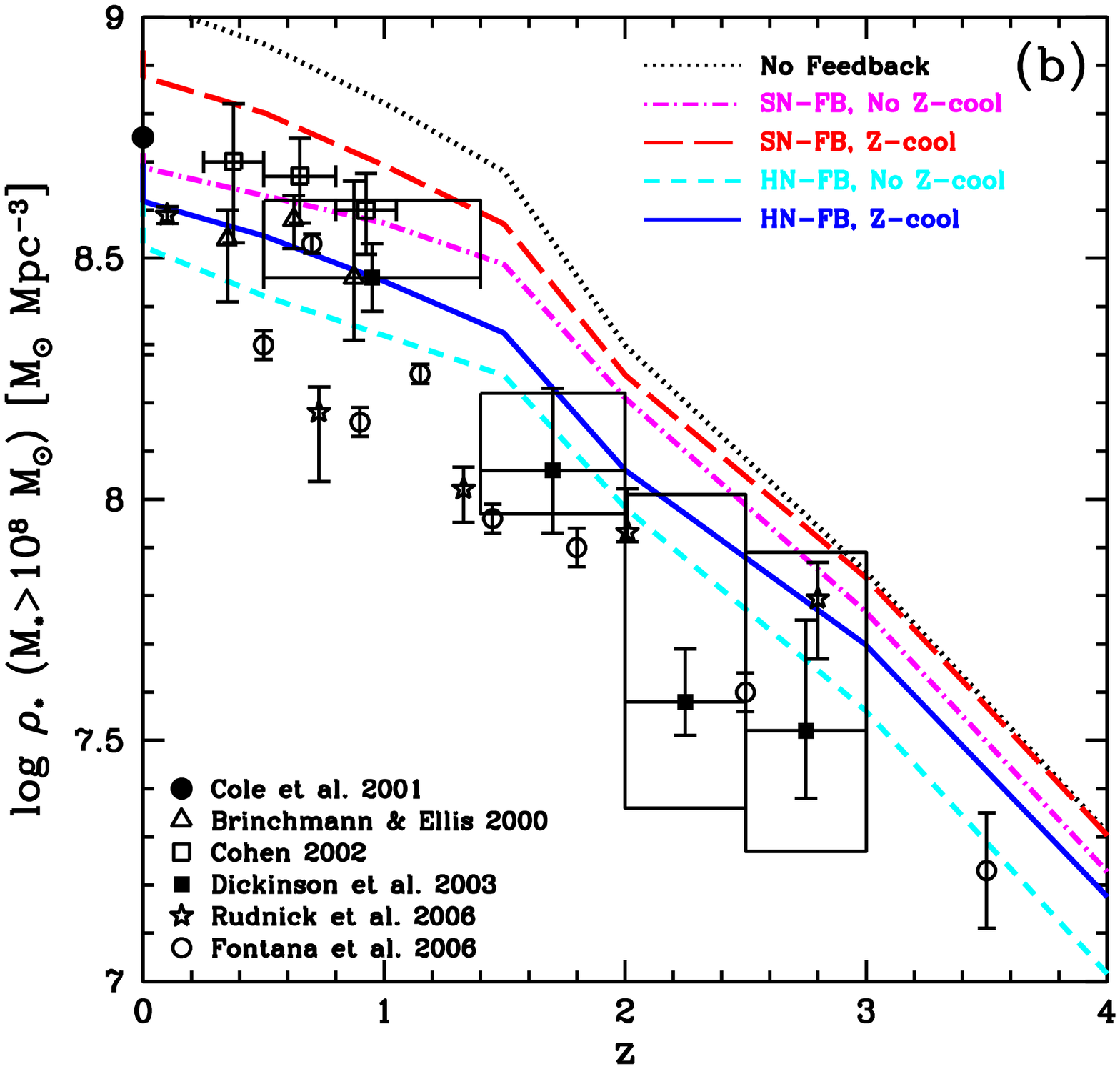}
\caption{\label{fig:sfr}
(a) Cosmic star formation rates for SN feedback with (dashed line) and without
  (dash-dotted line) metal-dependent cooling, and for HN feedback with (solid
  line) and without (dashed line) it.
  Rest-frame UV observations with dust correction are: 
  \citet{lil96}, open circles; \citet{mad96}, open squares;
  \citet{con97}, filled circles; \citet{ste99}, crosses; 
  \citet{bou03a}, filled triangles; \citet{iwa03}, filled pentagon; 
  \citet{bun04}, filled hexagon; \citet{gia04}, open triangles; 
  \citet{ouc04}, filled squares; \citet{sch05}, diamonds.
  H$\alpha$ observations are: \citet{gal95}, filled diamond; 
  \citet{per03}, \citet{gro99}, \citet{bri04}, \citet{tres98}, \citet{tre02} 
  increasing redshift with dust correction, filled stars.
  Submillimeter, radio and X-ray observations from: 
  \citet{hug98}, open star;
  \citet{bar00}, four-pointed stars;
  \citet{nor04}, asterisks.
(b) Same as panel (a) but for the stellar density evolution.
  Observational data sources are: \citet{col01}, filled circle;
  \citet{bri00}, triangles; \citet{coh02}, open squares; 
  \citet{dic03}, filled squares with the error boxes; 
  \citet{rud06}, stars; \citet{fon06}, open circles. 
}
\end{center}
\end{figure}

In Figure \ref{fig:sfr}a, we show how feedback and metal-dependent 
cooling affect the cosmic star formation history.
The cosmic SFRs that are directly measured from the
ages of stellar particles for the low resolution simulations ($N=56^3, N_{\rm
  FB}=72$).  Supernova feedback (dot-dashed line) decreases the SFR from
$z\sim3$, and the SFR is smaller by factor of $2$ at $0\ltsim z \ltsim2$ than
in the no feedback case (dotted line).  However, metal-dependent cooling
(long-dashed line) increases the SFR back to the no feedback case. For
comparison, Figure \ref{fig:sfr}a also shows observational estimates of the
cosmic SFR density at different epochs. Note that these determinations are
derived from the observed luminosity densities (e.g., \cite{mad96}), and this
involves uncertainties from dust extinction and completeness, as well as from
the IMF.
Rest-frame UV observations are plotted with dust correction with a factor of $2.7$ for $z<2$ and $4.7$ for $z>2$ (\cite{ste99}, \cite{red04}), and dust corrected H$\alpha$ measurements are taken from \citet{sch05}.

If we include hypernova feedback (dashed line), the SFR starts to be
suppressed from $z\sim6$ onwards, and is overall smaller by a factor of $3$ at
$0\ltsim z \ltsim3$.  The resulting SFR is in broad agreement with the
observations that show a peak of $\log({\rm SFR}/[{\rm M}_\odot{\rm yr}^{-1}])
\sim -1$ at $z\sim3$.  Since a hypernova releases ten times more energy than
an ordinary supernova, the distributed energy in one timestep ($\sim 10^{6-7}$ yr) is as large as
$\sim 10^{56}$ erg, and the temperature increase of individual gas particles
reaches $\sim 10^6$ K.  Once gas particles are heated to this temperature, the
gas particles do not cool rapidly due to the comparatively low cooling rate at
this temperature. This reduces not only the SFR but also the metal production
rate, even though the ejected metal mass from a single HN is larger than from
a SN.  The redshift where the mean gas [Fe/H] reaches $\sim -2$ is $z\sim3$
for the SN feedback, but is delayed to $z\sim2$ for the HN feedback.  Since
the metallicity dependence of the cooling rate clearly appears for [Fe/H]
$\gtsim -2$, metal-dependent cooling does not play such a prominent role in
the HN feedback case, and enriched gas can remain hot without forming stars.

In Figure \ref{fig:sfr}b, we show the time evolution of the stellar mass
density. To avoid uncertainties in the completeness from the faint end,
we measure the stellar mass in the galaxies with $M_* > 10^8 M_\odot$,
as in \citet{fon06}. 
The HN feedback improves the agreement with these observations.
The luminosity-limited observations (\cite{rud03,rud06}) are
also reproduced with the luminosity-limited measurements in our simulations.

The present stellar fraction (relative to total baryon density) and mean
metallicities are tabulated in Table~\ref{tab:cosmic}.  Without feedback,
$25\%$ of baryons turn into stars, which is too large compared with
observational estimates (e.g., \cite{fuk04}).  With SN feedback, the stellar
fraction reduces to $10-15\%$, which may be consistent with observation.
Recently, the observational estimate has been reduced to less than $10\%$
(\cite{fuk04}), which may require larger feedback.  The larger energy ejection
by HNe could provide a solution.  The present mean gas metallicity is [Fe/H]
$\sim -1.3$ for SN feedback, and is reduced to $\sim -1.8$ for HN feedback.
The mean stellar metallicity is almost solar for the SN feedback, and becomes
sub-solar for the HN feedback.

The feedback parameter $N_{\rm FB}$ does only weakly influence the cosmic SFRs.
With $N_{\rm FB}=36$, the SFR increases by a factor of $2$, which is in the
opposite sense from the isolated disk case (Section 3).  With the resolution
of our cosmological simulations, the ejected energy from one star particle is
large, and smaller $N_{\rm FB}$ results in weaker feedback due to a smaller
feedback radius.  For the higher resolution, because more small dense regions
are resolved, the SFRs at $z\gtsim4$ become much larger and the SFR at
$z\sim3$ increases by a factor of $3$.  To obtain similar feedback radius, we
adopt $N_{\rm FB}=405$ for the $N=96^3$ case, which gives a similar peak SFR
and stellar fraction to the low resolution case.  

We now discuss the dependence of our results on the star formation timescale
parameter $c_*$.  In the one-zone model, the star formation timescale directly
affects the star formation history, and it may be imagined that a smaller
$c_*$ can easily suppress the SFR.  However, in our chemodynamical model, the
influence of $c_*$ is more subtle than that.  With a smaller value for $c_*$,
say $c_*=0.02$, the peak redshift of the SFR is shifted towards lower redshift
($z\sim3$) and the present SFR density is enlarged ($\log{\rm SFR}=-1.4$), but
the total amount of stars is not reduced in any significant way.  The reason
is that if we prevent the gas from forming stars at some early epoch, this
will only delay the transformation of cold gas into stars, but the amount of
gas that cools and becomes available for star formation is not changed. If we
pick a value as large as $c_*=1$, the initial star burst is so large that the
SFR continuously decreases from $z=6$.  Comparing with the observed cosmic
SFRs, we therefore adopt $c_*=0.1$, which roughly reproduces the observed
shape of the cosmic SFR history, keeping in mind however that the latter is
still quite uncertain due to dust extinction, for example.  If the SFR at
$z\gtsim3$ is as low as the UV observations without a dust correction suggest,
and if the SFR at $z\ltsim1$ is as high as the IR observations with the dust
correction suggest, then a value for $c_*$ as small as $0.02$ may be allowed.
We also note that the $c_*$ parameter changes the size of galaxies, and it
can hence be constrained by comparing simulation results with the
observed scaling relations of galaxies (see \cite{kob05} for the detail).

\begin{figure}
\begin{center}
\includegraphics[width=8cm]{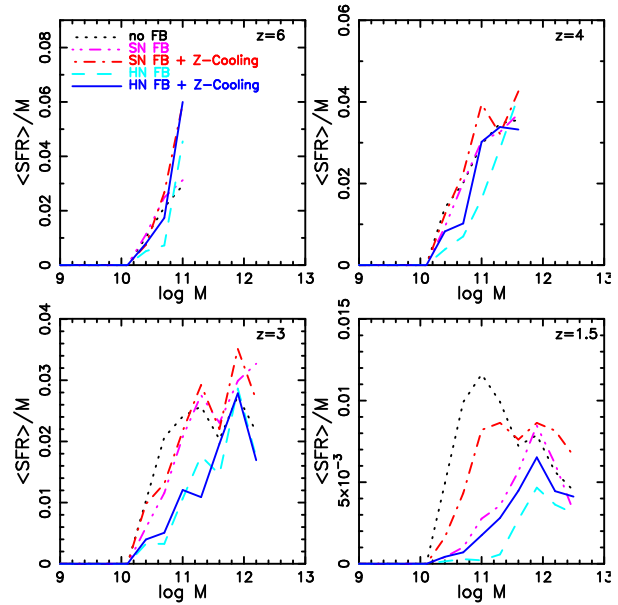}
\caption{\label{fig:galsfr}
  The average SFR normalized of the total mass for the models without feedback
  (dotted line), with SN feedback without (dash-dot-dot-dot line) and with
  (dot-dashed line) metal-dependent cooling. In addition, results are shown
  for HN feedback without (dashed line) and with (solid line) metal-dependent
  cooling.  }
\end{center}
\end{figure}

To investigate where the SFR is suppressed with HN feedback, we show the
average SFRs at given total mass in Figure \ref{fig:sfr}.  At $z=6$, the SFRs
in massive galaxies are enhanced with the metal-dependent cooling (dot-dashed
and solid lines).  After $z=3$, the SFRs in low mass galaxies are suppressed
by SN feedback (dot-dashed and dashed-dot-dot-dotted lines), but are
unaffected in massive galaxies.  If we include HN feedback (solid and dashed
lines), the SFRs are smaller for all galaxies, independent of metal-dependent
cooling.  The stellar mass of the most massive galaxies are not altered
significantly ($8.5$, $6.1$, $7.0$, $5.8$, $5.0 \times 10^{11}M_\odot$
respectively for the models in the label).  However, the number of galaxies
with $M_*\sim 10^{9-11}{\rm M}_\odot$ decreases by a factor of $2-3$, and the
number of dwarf galaxies with $M_*\ltsim 10^{9}{\rm M}_\odot$ increases by a
factor of $2-3$.

\begin{figure}
\begin{center}
\includegraphics[width=12.5cm]{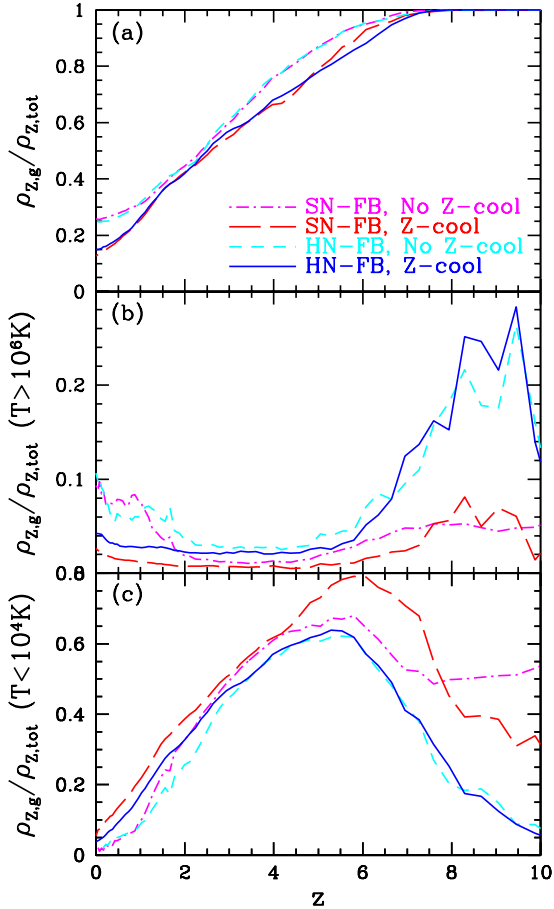}
\caption{\label{fig:metal}
  The redshift evolution of the metal fraction, normalized by the total
  projected metals, in (a) all the gas, (b) hot ($T>10^6\,{\rm K}$) gas, and
  (c) cold ($T< 1.5 \times 10^4\,{\rm K}$) gas.  See Fig.~\ref{fig:sfr} for
  the models.  }
\end{center}
\end{figure}

\begin{figure}
\begin{center}
\includegraphics[width=12.5cm]{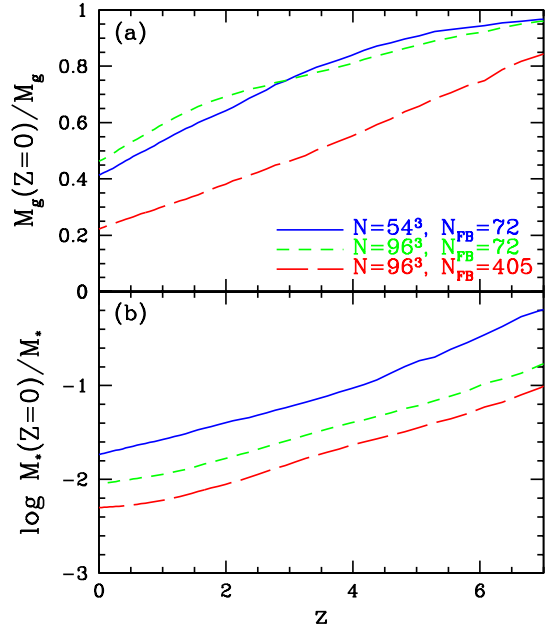}
\caption{\label{fig:zerometal}
  The fractions of metal-free (a) gas and (b) stars.  The solid, short-dashed,
  and long-dashed lines are for the models with HN feedback and metal-dependent
  cooling, respectively, with $(N=54^3, N_{\rm FB}=72)$, $(N=96^3, N_{\rm
    FB}=72)$, and $(N=96^3, N_{\rm FB}=405)$.  }
\end{center}
\end{figure}

The feedback and the metal-dependent cooling affect the distributions of heavy
elements.  Figure~\ref{fig:metal} shows the redshift evolution of the metal
fraction, normalized by the total projected metals, in all gas, in hot gas
($T>10^6$ K), and in cold ($T< 1.5 \times 10^4$ K) gas.  Because of star
formation, the metal fraction increases in stars (panel a) and decreases in
gas towards lower redshifts.  At the beginning, metals are ejected into hot
gas (panel b), which eventually cools and increases the metal fraction in cold
gas toward $z\sim5$ (panel c).  From $z\sim5$ onward, the metal fraction in
cold gas decreases due to star formation.  At $z\sim3$, almost half of all
metals are in stars.  With the metal-dependent cooling (long-dashed and solid
lines), more metals are locked into stars than is the case without it
(dot-dashed and short-dashed lines).  With HN feedback (short-dashed and solid
lines), two times more metals exist in gas, more in hot gas and less in cold
gas.  This metal fraction in hot gas, however, is still smaller than expected
from observations of the X-ray gas in clusters (larger than two-thirds,
\cite{ren02}).
\citet{dav06} showed the increase of the metal fraction in hot gas
toward low-redshifts using a hydrodynamical model with a momentum
driven wind.  Such increase is not seen in our simulations with
metal-dependent cooling.

Figure~\ref{fig:zerometal} shows the redshift evolution of the fractions of
metal-free gas and metal-free stars.  In our model with HN feedback and
metal-dependent cooling (solid line), half of the gas remains pristine until
$z\sim3$, and $\sim 20\%$ gas is still metal-free at $z=0$ (panel a).  The
fraction of metal free stars is as large as $\sim 50\%$ at $z\sim6$ and
decreases to $\sim 5\%$ at $z=0$ (panel b).  Higher resolution (short-dashed
line) gives almost the
same fraction of metal-free gas. However, the amount of metal-free gas
strongly depends on the number of feedback neighbours $N_{\rm FB}$ in our
model. A larger value for $N_{\rm FB}$ gives a smaller fraction of metal-free
gas (long-dashed line).  We should note that we do not include an explicit
mixing of metals between neighbouring gas particles.  Gas particles are
enriched only when they pass near dying star particles in our model.  Although
an additional mixing effect could increase the enriched fraction of gas, we
note that the region enriched by a supernova is already large in our
simulation model.

\begin{figure}
\begin{center}
\includegraphics[width=7.5cm,angle=-90]{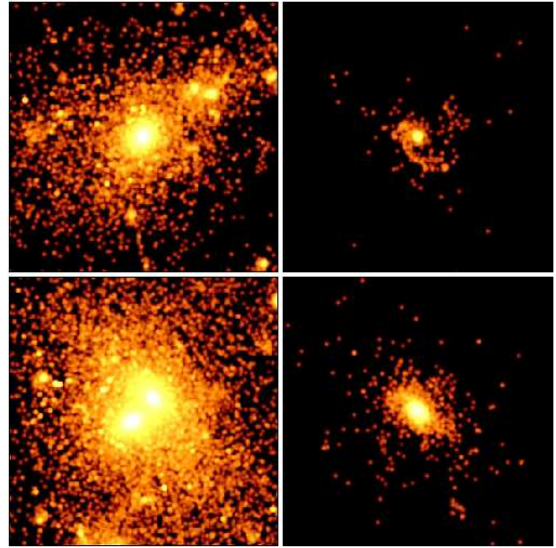}
\caption{\label{fig:galmap}
  V-band images of selected galaxies in our cosmological simulation at $z=0$.
  The panels are $200$ kpc on a side, and the galaxy images are ordered by
  their total mass.  }
\end{center}
\end{figure}

\subsection{Star Formation, galactic winds, and chemical enrichment}

\begin{figure}
\begin{center}
\includegraphics[width=6.1cm,angle=-90]{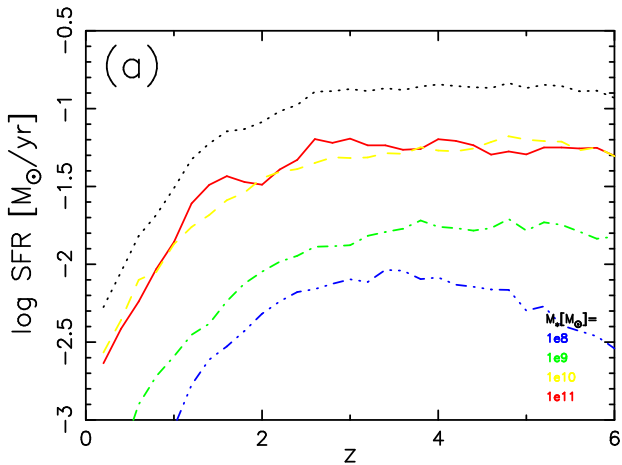}
\caption{\label{fig:sfr_deco}
  The cosmic SFR history split by different galaxy mass scales at $z=0$. We
  show results for stellar masses of $10^{11}$ (solid line), $10^{10}$ (dashed
  line), $10^9$ (dot-dashed line), and $10^8 M_\odot$ (dot-dot-dot-dashed
  line). The dotted line shows the total.  The galaxies are identified by FOF
  at $z=0$, so these SFRs correspond to the stellar age distributions in the
  galaxies.  }
\end{center}
\end{figure}

\begin{figure}
\begin{center}
\includegraphics[width=6.1cm,angle=-90]{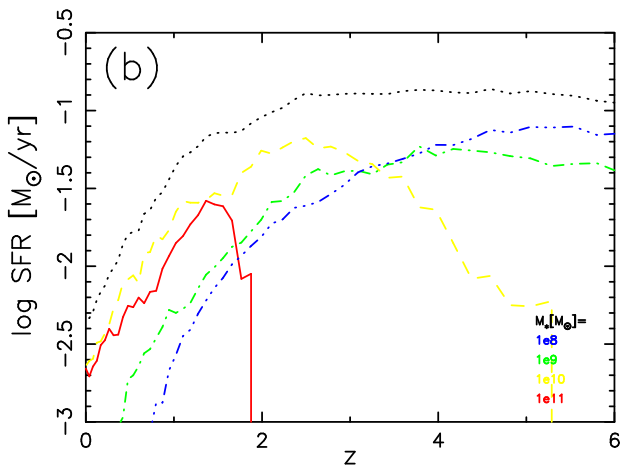}
\caption{\label{fig:sfr_deco2}
  The same as Fig.~\ref{fig:sfr_deco}, but the galaxies are here identified at
  each redshift, so the mass quoted corresponds to the current one at each
  redshift.  }
\end{center}
\end{figure}

In the following, we discuss in more detail the results of the higher
resolution simulation ($N=96^3, N_{\rm FB}=405$) with HN feedback and
metal-dependent cooling.  Figure~\ref{fig:galmap} shows V-band images for 
our galaxies identified by the FOF algorithm at $z=0$.
The most massive galaxy shown (upper left panel) is a merging system, with the total mass $M_{\rm tot} \sim 1.5\times10^{13} {\rm M}_\odot$ and the stellar mass $M_* \sim 4\times10^{11} {\rm M}_\odot$.
The others illustrate the typical morphologies of our simulated galaxies with $M_{\rm tot} \sim 6\times10^{12}, 8\times10^{11}$, and $2\times10^{11} {\rm M}_\odot$, and $M_* \sim 1\times10^{11}, 4\times10^{10}$, and $4\times10^{9} {\rm M}_\odot$, respectively.

When and where do stars form? To answer this question,
we break up the cosmic SFR history according to galaxy mass. In
Figure~\ref{fig:sfr_deco}, we show the SFR history of systems
with final stellar masses of $M_* \sim 10^{11}$ (solid line), $10^{10}$ (dashed line),
$10^9$ (dot-dashed line), and $10^8 {\rm M}_\odot$ (dot-dot-dot-dashed line).
The dotted line shows the total, corresponding to the $\sim1000$ galaxies in
this simulation. The galaxies have been identified by FOF at $z=0$ for this
plot, and thus these SFRs correspond to the age distribution of stars in the
galaxies.  For all galaxy masses, the SFRs show a peak around $z\sim3-4$, and
the majority of stars are as old as $\sim 10$ Gyr.  On the other hand, in
Figure~\ref{fig:sfr_deco2}, we identify galaxies at each redshift and split up
the SFRs according to the current stellar mass measured at the redshift,
which are comparable to the observations of high redshift galaxies.
This shows that most stars have formed in low-mass galaxies
with $10^{8-9}{\rm M}_\odot$ at high redshift $z \gtsim 3$. 
$10^{10}M_\odot$ galaxies exist at high redshift $z \ltsim 5$, but $10^{11}M_\odot$ galaxies appear only after $z \sim 2$.
From these two
figures we conclude that most stars have formed in dwarf galaxies 
before they merge to massive galaxies in our simulation.
As a result of the
hierarchical clustering of dark matter halos, such old stars belong to massive galaxies at low redshifts.

If we consider luminosity-weighted ages, the largest galaxies need not always
be the oldest if young stars keep on forming in them.  At the present epoch,
we find no relation between the luminosity-weighted ages and the stellar mass
of the galaxies, and most galaxies are as old as $\sim 10\,{\rm Gyr}$.  Large
galaxies with $M_* \gtsim 10^{11}\,{\rm M}_\odot$ tend to have young ages as $\sim 8$ Gyr
in our model.  In these galaxies, feedback from active galactic nuclei (AGN)
may stop recent star formation.  For dwarf galaxies, the ages span a wide
range of $1-10$ Gyr, and there exist two populations, young and old dwarfs
that correspond to the observed dwarf irregulars and spheroidals,
respectively.  However, the number of young dwarfs seems to be too small
compared with observations (e.g., \cite{kau03}), and the so-called
`down-sizing' effect of cosmic star formation is not clearly seen in our
simulations.  This is because i) star formation is not terminated in massive
galaxies, presumably due to the lack of efficient feedback and ii) not enough
young dwarfs are forming at $z=0$.  AGN feedback (e.g., \cite{dim05},
\cite{cro06}) may solve the former problem, but not the latter.

\begin{figure*}
\begin{center}
\includegraphics[width=7.5cm,angle=-90]{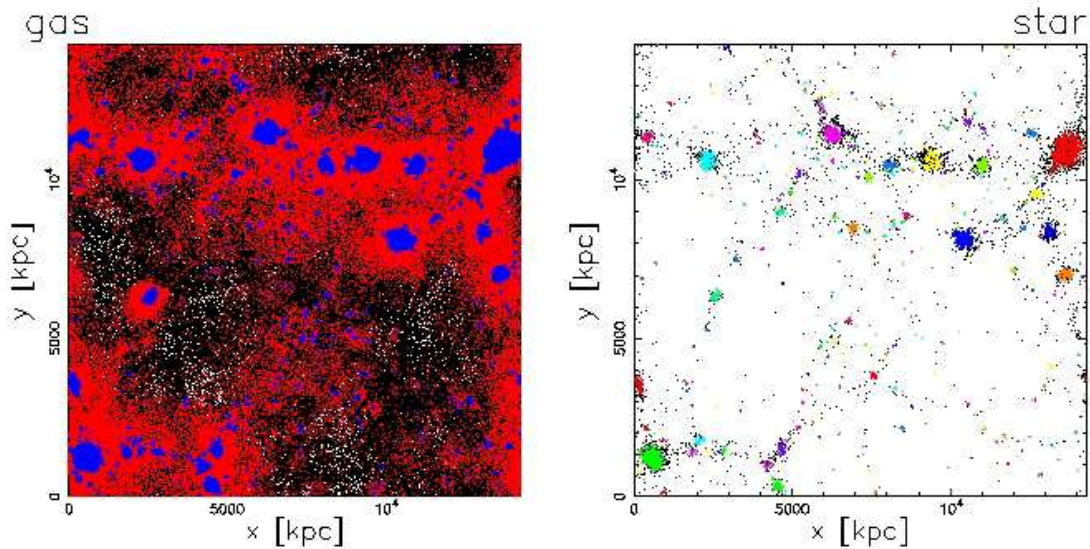}
\caption{\label{fig:windmap}
  Spatial distribution of wind particles at $z=0$. The blue and red points
  show gas particles in galaxies and in the wind phase, respectively.  The
  distribution of star particles is given in the right panel, with different
  colours for different galaxies. }
\end{center}
\end{figure*}

\begin{figure*}
\begin{center}
\includegraphics[width=14.cm]{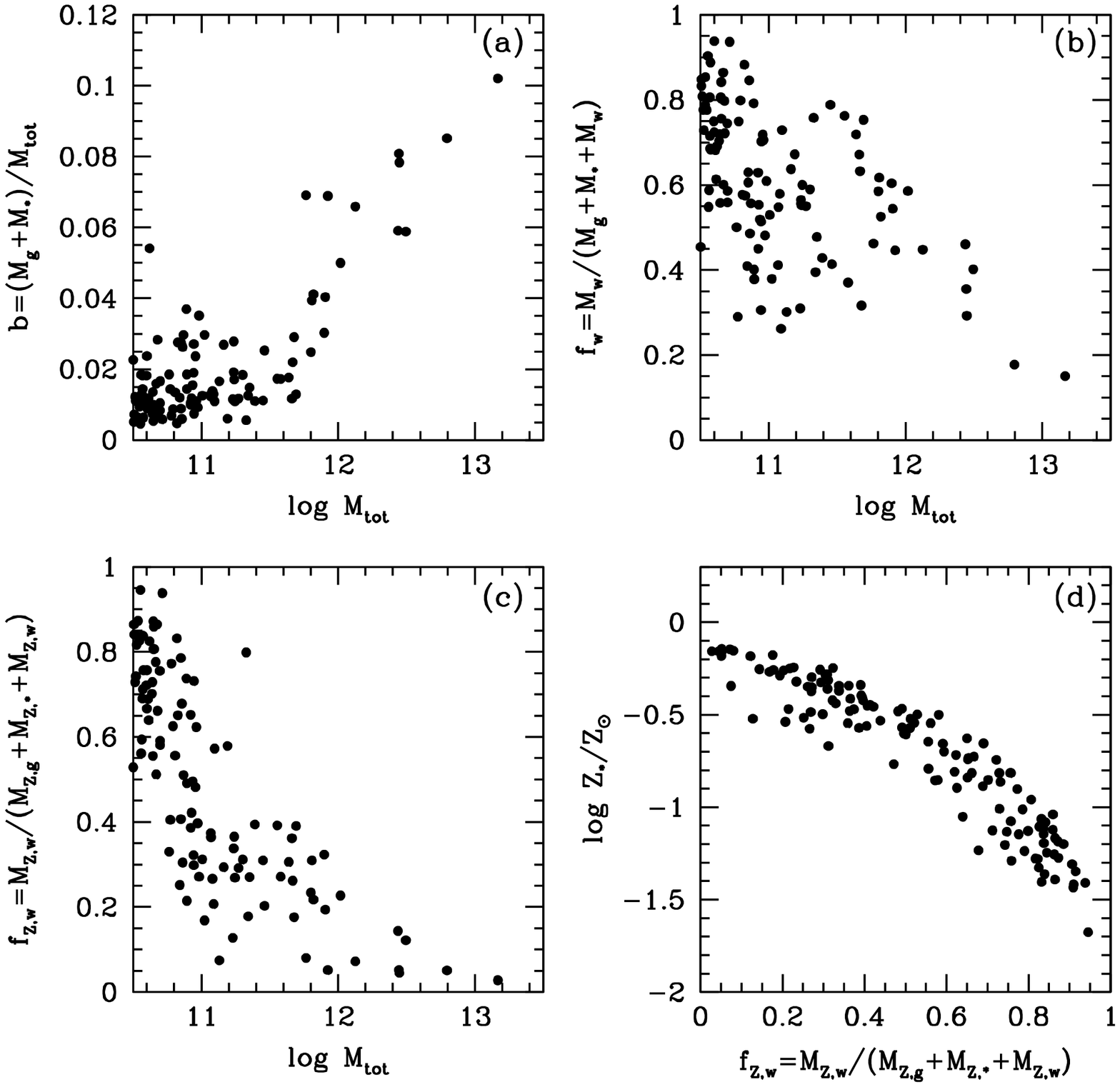}
\caption{\label{fig:wind}
  Baryon fraction (panel a), wind fraction $f_{\rm w}=M_{\rm w}/(M_{\rm g}+M_*+M_{\rm w})$ (panel b), and ejected metal fraction $f_{\rm w,Z}=M_{\rm Z,w}/(M_{\rm Z,g}+M_{\rm Z,*}+M_{\rm Z,w})$ (panel c) against total mass. Panel (d) shows the stellar metallicity against the ejected metal fraction.  }
\end{center}
\end{figure*}

How are heavy elements ejected from galaxies to the IGM?  In the simulation,
we can trace the orbit of gas particles over time. Exploiting this, we define
as wind particles those that are not in galaxies now, but have been in
galaxies before.  Figure~\ref{fig:windmap} shows the spatial distributions of
this `wind gas'.  The blue and red points mark the positions of gas particles
in galaxies and of gas in the wind, respectively.  The corresponding
distribution of star particles is shown in the right panel.  The wind
particles surround the stars in galaxies, but do not extend very far from
them.  In this simulation, $\sim 10\%$ of baryons turn into stars, $\sim 10\%$
of the gas stays in galaxies ($\sim 8\%$ is hot), and $\sim 20\%$ is ejected
as galactic winds (see Table~\ref{tab:cosmic} for the other
models).  The rest, half of the baryons, never accretes onto galaxies.

\begin{figure}
\begin{center}
\includegraphics[width=6.5cm]{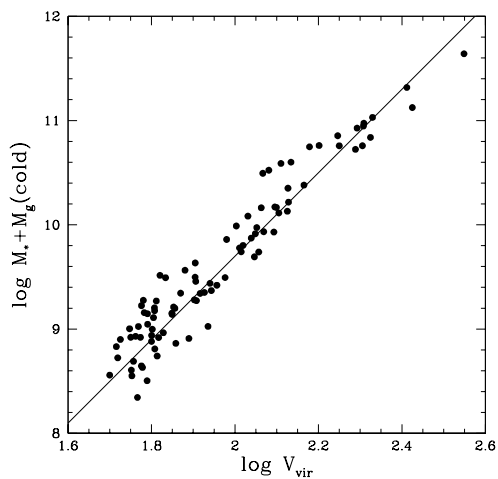}
\caption{\label{fig:tully}
  Baryonic Tully-Fisher relation.
The solid line shows the observation taken from \citet{mcg05}.}
\end{center}
\end{figure}

When we follow the orbits of gas particles, we can also examine from which
galaxies the wind gas particles are ejected. This allows a measurement of
the ejected wind mass from each galaxy.  It is possible that a wind gas
particle is ejected from two different galaxies (at different times), in which
case we count it for both galaxies. But if a wind gas particle should be
ejected twice from the same galaxy, it is only counted once.  In
Figure~\ref{fig:wind}, we plot the wind fraction, i.e.~the ratio between the
total wind mass to the total accreted baryon mass, against the total mass.  Here we
plot only galaxies with $M_* > 10^8\,{\rm M}_\odot$ because
of the limited resolution.  Massive galaxies are dominated by baryons
($b\sim0.1$), but the baryon fraction is smaller for lower-mass galaxies
(panel a), which implies that baryons are more efficiently ejected from them. 
Most of the gas in our massive galaxies is in the hot phase. Therefore,
the difference in the baryon fraction does not conflict with the
observed baryonic Tully-Fisher relation (\cite{mcg05,pfe05}), 
as we explicitly show in Figure~\ref{fig:tully}.

A clear relation is found between the wind fraction
and the total mass (panel b).  Winds are efficiently ejected from small
galaxies, with $\sim 80\%$ of accreted baryons being ejected from $M_{\rm
  tot}\sim 10^{11}{\rm M}_\odot$ galaxies.
A similar relation is also found for the ejected metal fraction, i.e.~the ratio between the wind metal mass to the total metal mass (panel c).
It is interesting that the wind
fraction and the ejected metal fraction correlate well with the stellar metallicity (panel~d).  Based on
this finding, we conclude that the origin of the mass-metallicity relation can
be explained with galactic winds.

We note that the mass contribution of the winds is still larger for
massive galaxies. $\sim 50\%$ of the total wind mass and $\sim 40\%$
of the total metals are ejected from massive galaxies with $M_{\rm
tot} > 10^{12}\,{\rm M}_\odot$.  The smaller contribution for the metal
ejection from massive galaxies is due to the metal-dependent cooling;
enriched gas tends to be locked in stellar population of galaxies.

\begin{figure}
\begin{center}
\includegraphics[width=8.5cm]{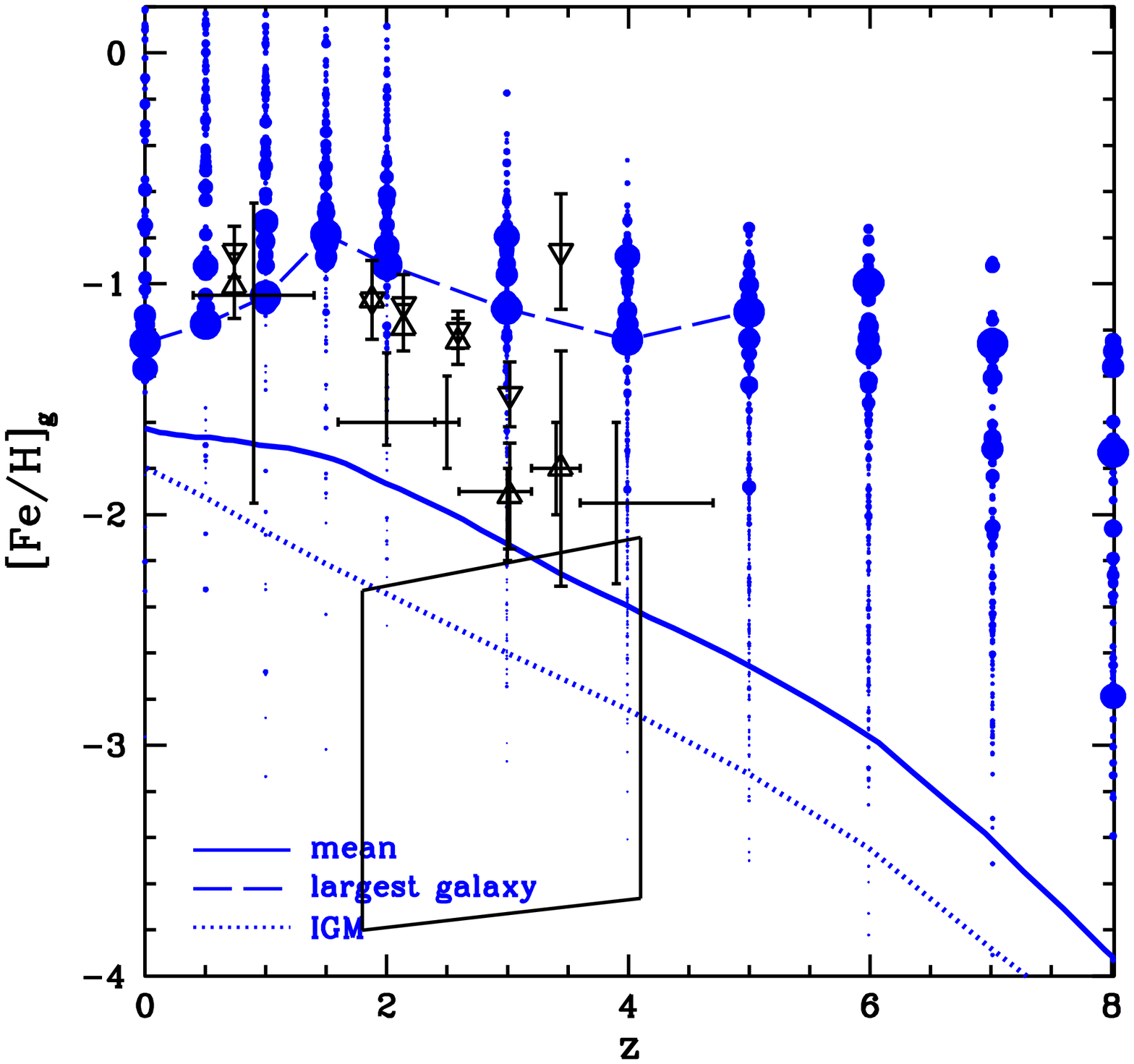}
\includegraphics[width=8.5cm]{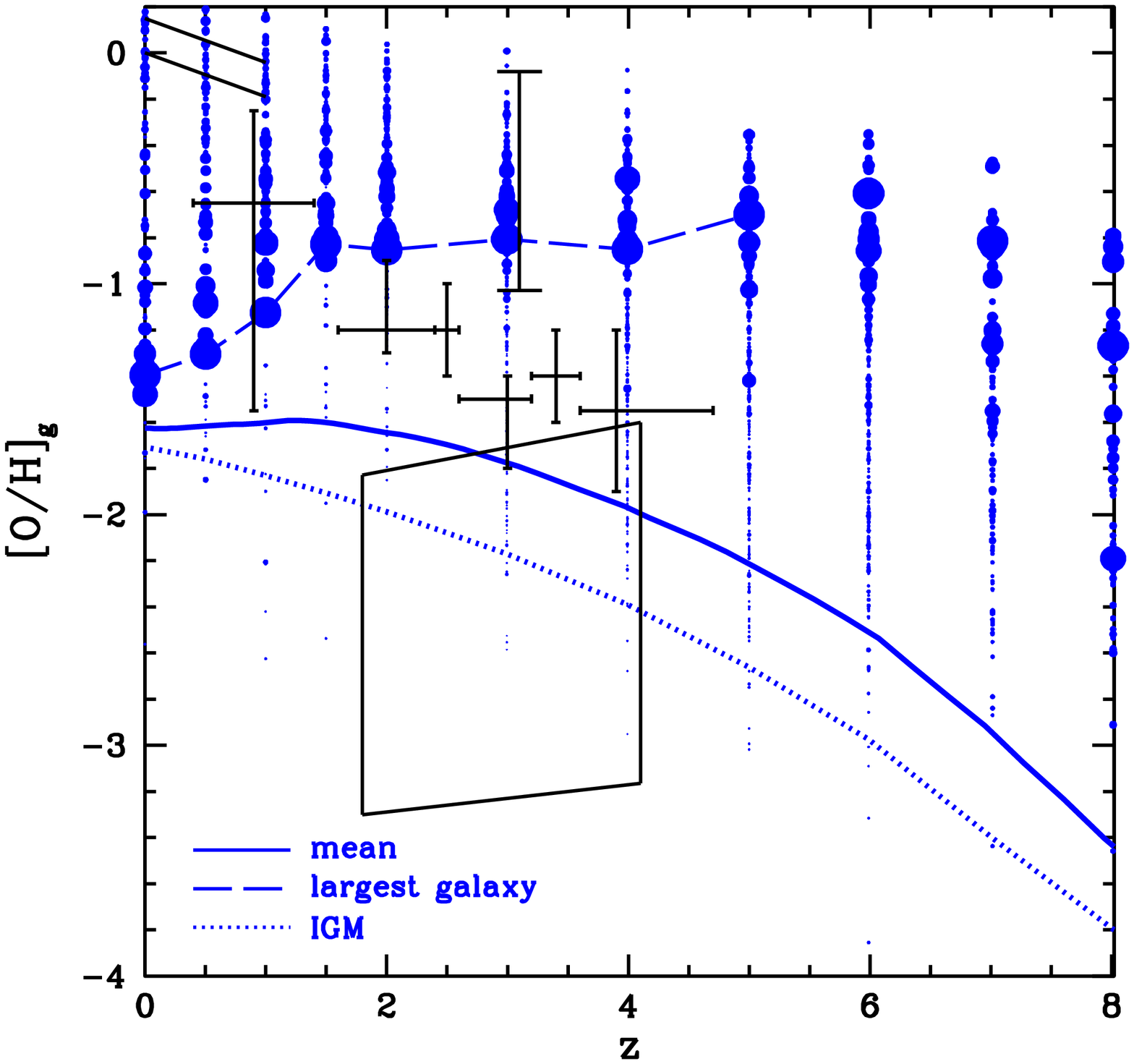}
\caption{\label{fig:feh}
  Redshift evolution of iron and oxygen abundances of the gas.  The points
  show individual galaxies with the size of the symbols encoding the size of
  the galaxies.  The solid, dashed, and dotted lines are for the mean, the
  largest galaxy, and the IGM, respectively.  The observational data are:
  The large errorbar at $z=3$ for the LBGs of \citet{pet01}, 
  errorbars for the DLA systems of \citet{pro03} where [O/Fe]=0.4 adopted,
  triangles for the DLA systems of \citet{kul05} where [Zn/Fe]=0 adopted,
  and the two parallel lines at $z=0-1$ for the
  emission-line galaxies of \citet{kobu04}.  The box is for [C/H] in the IGM
  (\cite{sch03}), shifted with our assumption of [C/Fe]$=-0.5$ and [C/O]$=-1$.
}
\end{center}
\end{figure}

The metal enrichment timescale depends on the environment.  Figure
\ref{fig:feh} shows the evolution of iron and oxygen abundances in the gas
phase.  The average metallicity of the universe (solid line) reaches [O/H]
$\sim -2$ and [Fe/H] $\sim -2.5$ at $z\sim4$ ($1.5$ Gyr), but reaches the same values at $z\sim3$ ($2.1$ Gyr) in the IGM (dotted line).  The box
shows [C/H] obtained from quasar absorption line measurements, which involves
some uncertainty in the UV background radiation.  [C/Fe] is also uncertain,
and spans a wide range of $-0.6 \ltsim$ [C/Fe] $\ltsim 1$ in the metal-poor
stars in the Milky Way.  Since core-collapse supernovae cannot produce so much
carbon relative to Fe ([C/Fe] $\sim -0.4$, \cite{kobnom06}), the C-rich stars
seem to be enriched by massive and/or low-mass stellar winds, they could also
be affected by internal mixing or by external enrichment from a binary
companion.  The metallicity of the IGM seems not to be affected significantly
by stellar winds because of the lower energy and the different timescale.
Assuming that the IGM is enriched only by core-collapse supernovae, we could
account for this by shifting the observed region assuming [C/Fe]$=-0.5$ and
[C/O]$=-1$, and then our result becomes comparable to the observations.  We
note that this assumption can explain the observed [Si/C] in the IGM
(\cite{agu04}), while pair instability supernovae, which produce much more
iron and less carbon, cannot resolve this conflict.  A self-consistent model
for the carbon abundance is required.

The metallicities of the individual galaxies with $M_{\rm g}+M_*>10^{9}{\rm
  M}_\odot$ are shown as points, with larger points representing more massive
galaxies.  In large galaxies, the enrichment takes place so quickly that [O/H]
reaches $\sim -1$ at $z\sim7$.  This is consistent with the metallicities of
LBGs (large errorbar at $z=3$, \cite{pet01}).  Smaller galaxies evolve with
more variable timescales, resulting in a larger scatter of $-2 \ltsim$ [O/H]
$\ltsim 0$ at present.  The metallicities of DLA systems (errorbars,
\cite{pro03}; triangles, \cite{kul05}) are comparable with our galaxies, 
suggesting an identification
of DLAs with dwarf galaxies or the outskirts of massive galaxies.

The metallicities measured for local emission-line galaxies (two parallel
lines at $z=0-1$, \cite{kobu04}) are much larger than our simulation
predictions.  One reason could be an aperture effect.  The metallicities
of our model galaxies are measured for the whole galaxy, which may be
systematically low compared to the observations since strong metallicity
gradients are found in our galaxies (Fig.~\ref{fig:cosmicmap}) and the
observational data focuses on star-forming regions which may be biased towards
galaxy centres.
The other reason is that a lot of metal-poor hot/warm gas exists 
in our large galaxies because of their deep potential.
Observing the metallicity of warm gas is also important to put constraint on the chemical enrichment history of the universe.
If we measure the metallicity of the cold gas in the innermost $10$ kpc 
region, [O/H] reaches $\sim 0.2$, weakly increasing toward $z=0$, which is comparable to the observations.

\begin{figure}
\begin{center}
\includegraphics[width=8.5cm]{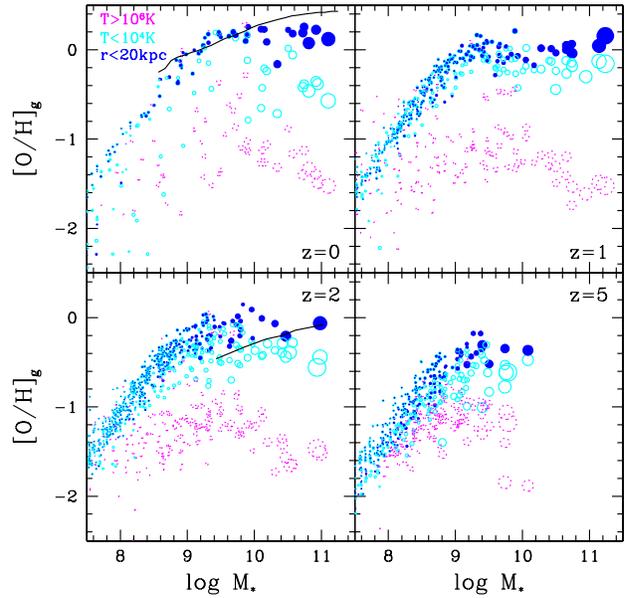}
\caption{\label{fig:feh_mass}
  Mean metallicities of hot ($T>10^6$ K, dotted-open circles) and cold gas
  ($T<10^4$ K, open circles), plotted against the total stellar mass.  The filled
  circles are for the cold gas within 10 kpc.  The solid lines show the
  relation for the nebular gas observed with the emission lines for SDSS
  galaxies (\cite{tre04}) at $z=0$ and star-forming galaxies (\cite{erb06}) at $z=2$.}
\end{center}
\end{figure}

\begin{figure}
\begin{center}
\includegraphics[width=8.5cm]{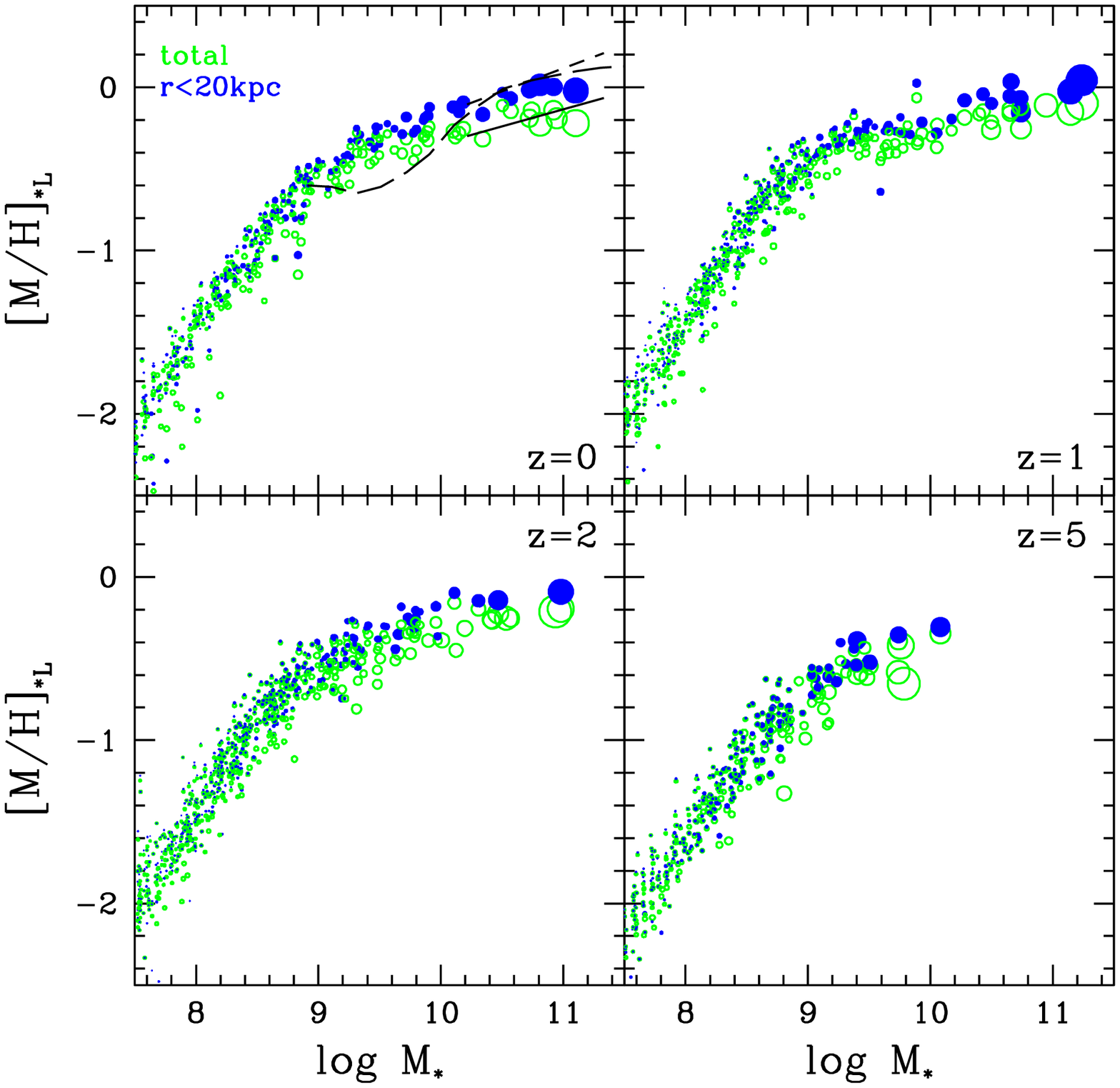}
\caption{\label{fig:fehs_mass}
  Relation between the V-band luminosity-weighted stellar metallicity and the 
  total stellar mass.
  The open and filled circles are for the whole galaxy and central region within 10 kpc, respectively.
The solid and short-dashed lines are for the observed
  relations in nearby ellipticals with measured Mg$_2$ index, the average is
  derived accounting for the metallicity gradients (\cite{kob99}) and the
  central values (\cite{pah98}), respectively.  The long-dashed line is for
  SDSS galaxies (\cite{gal05}).  }
\end{center}
\end{figure}

In Figure~\ref{fig:feh_mass},
the metallicities of hot (dotted-open circles) and cold (open
circles) gas are shown against the stellar mass of the galaxies.
There seems to be a relation that gives a larger
metallicity of the cold gas in more massive galaxies, and [O/H] $\sim 0$ and
$-1.5$ in giant and dwarf galaxies, respectively.  The filled circles show the
cold gas metallicities within $10$ kpc, which are comparable to those found in
SDSS emission-line galaxies (solid line, \cite{tre04}), but the slope is
shallower.  In our model, mass-metallicity relations for central cold gas are
found after $z\sim5$, although there is a significant scatter.
At $z=2$, the gas metallicity is consistent with the observation (solid
line, \cite{erb06}) for massive galaxies, but seem to be larger than
observed for low-mass galaxies.

In contrast to this large scatter, tight relations are found between the
stellar metallicity and stellar mass at any redshift, as shown in
Figure~\ref{fig:fehs_mass}, based on V-band luminosity-weighted metallicities.
The mean stellar metallicity of the whole galaxy (open circles) reaches solar for massive
galaxies, and half solar for intermediate-mass galaxies.  This is consistent
with observations of nearby ellipticals when the effect of metallicity
gradients is taken into account (solid line, \cite{kob99}).  The central
metallicity measured within $10$ kpc (filled circles) is two times higher (short-dashed line,
\cite{pah98}) than a global average.  In our simulation, progressively more
material has been blown away via galactic winds from present-day lower mass
galaxies (Fig.~\ref{fig:wind}), and we find a mass-metallicity relation that
continues for dwarf galaxies.  However, \citet{gal05} (long-dashed line) found
a change of the slope at $M_* \sim 10^{10}{\rm M}_\odot$ in the SDSS, which
does not appear in our simulations.  The change of the slope at $M_* \sim 10^9
{\rm M}_\odot$ in our results is presumably due to resolution effects.

\section{Conclusions}

We have simulated cosmic chemical evolution with a tree-SPH code including
star formation, supernova feedback, and a detailed chemical enrichment model
that does not rely on the instantaneous recycling approximation. We have also
investigated the role hypernova may play for regulating star formation and for
enriching the universe. Our main findings can be summarized as follows.

\begin{enumerate}
\item Supernova feedback can suppress star formation, but this reduction is
  largely compensated by an increase of similar size when metal-dependent
  cooling is included.  In simulations of isolated disk galaxies with a total
  halo mass of $10^{10}\,h^{-1} {\rm M}_\odot$, a bipolar galactic wind is formed as a
  result of the supernovae feedback and ejects up to half of the baryons.  The
  ejected metal fraction is not as large because of metal-dependent cooling,
  and amounts to a few percent for supernova feedback. If hypernovae are
  included as well, this raises to $25\%$. In this case, more metals are
  ejected with more energy, and the enriched gas is heated so strongly that
  rapid subsequent cooling in prevented.  The wind efficiency does not depend
  strongly on numerical resolution.  However, it depends strongly on halo
  mass; for massive halos of mass $10^{12}\,h^{-1} {\rm M}_\odot$ no winds occur.
  
\item The cosmic SFR shows a peak at $z\sim4$, with $\sim 10\%$ of baryons
  turning into stars, roughly consistent with observational estimates.  We
  analyse the contributions to the cosmic SFR from different galaxy mass
  scales at present (Fig.~\ref{fig:sfr_deco}), and at the observed redshift
  (Fig.~\ref{fig:sfr_deco2}).  The majority of the stars in present-day
  massive galaxies have formed in smaller systems at high redshifts, and the
  stellar population of these galaxies is as old as $10\,{\rm Gyr}$,
  consistent with observations of elliptical galaxies.  No relation is found
  between luminosity-weighted age and stellar mass.  The ages of low-mass
  galaxies span a wide range of $1-10$ Gyr.
  
\item Tracing the orbits of gas particles, we found that $\sim 20\%$ of all
  baryons are ejected at least once from galaxies into the IGM.  Galactic
  winds blow particularly efficiently from low-mass galaxies.  The
  wind fraction is larger for less massive galaxies, and $\sim80\%$ of the
  accreted baryons are ejected from galaxies with mass below $M_{\rm
    tot}\sim10^{11} {\rm M}_\odot$ (Fig.~\ref{fig:wind}).  The wind fraction
  correlates well with the stellar metallicity, and the origin of the mass-metallicity relation is from galactic winds.
  
\item The metal-enrichment history depends on environment. In large galaxies,
  enrichment takes place so quickly that [O/H] reaches $\sim -1$ at $z\sim7$,
  which is consistent with the sub-solar metallicities of the Lyman break
  galaxies.  The low metallicities of DLA systems are also consistent with our
  galaxies, provided these systems are dwarf galaxies or the outskirts of
  massive galaxies.  The low [C/H] of the IGM can be explained if the IGM is
  enriched only by SNe II and HNe.  The average metallicity of the universe
  reaches [O/H] $\sim -2$ and [Fe/H] $\sim -2.5$ at $z\sim4$ ($1.5$ Gyr), but
  reaches the same values at $z\sim3$ ($2.1$ Gyr) in the IGM.
  
\item In galaxies, metallicity of the cold gas increases with galaxy mass, 
  which is comparable to observations with a large scatter.
  The central cold gas shows a relation between galaxy mass
  and metallicity with shallower slope than observed in emission-line
  galaxies.  For the stellar population, the observed mass-metallicity
  relation is well reproduced, and originates in mass-dependent galactic winds.
  These relations are present since $z \sim 5$.
\end{enumerate}

At $z=0$, however, star formation has not terminated in our massive galaxies,
and not many dwarf galaxies are still forming.  The latter problem may be due
to lack of resolution, while the former may reflect non-supernova feedback
such as accretion onto supermassive black holes. It will be interesting to
include active galactic nuclei in future studies of cosmic chemical evolution,
as well as to further increase the dynamic range of these calculations. This
will then allow methods such as the ones explored here to fully exploit the
observational abundance measurements to better constrain galaxy formation
physics.

\section*{Acknowledgements}
C.K. thank to the Japan Society for Promotion of Science for a financial support.

\bsp

\label{lastpage}


\begin{thebibliography}{}

\bibitem[\protect\citeauthoryear{Aguirre et al.}{2004}]{agu04}
Aguirre, A., Schaye, J., Kim, T.-S., Theuns, T., Rauch, M., Sargent, W. L. W. 2004, \apj, 602, 38

\bibitem[\protect\citeauthoryear{Arimoto \& Yoshii}{1987}]{ari87} 
Arimoto, N., \& Yoshii, Y. 1987, \aap, 173, 23

\bibitem[\protect\citeauthoryear{Arnaud et al.}{1992}]{arn92}
Arnaud, M., Rothenflug, R., Boulade, O., Vigroux, L., \& Vangioni-Flam, E. 1992, \aap, 254, 49

\bibitem[\protect\citeauthoryear{Barger, Cowie, \& Richards}{2000}]{bar00}
Barger, A. J., Cowie, L. L., \& Richards, E. A. 2000, \aj, 119, 2092

\bibitem[\protect\citeauthoryear{Bromm \& Larson}{2004}]{bro04}
Bromm, V. \& Larson, R. B. 2004, \araa, 42, 79

\bibitem[\protect\citeauthoryear{Bouwens, Broadhurst, \& Illingworth}{2003ab}]{bou03a}
Bouwens, R. J. Broadhurst, T. \& Illingworth, G. 2003, \apj, 593, 640

\bibitem[\protect\citeauthoryear{Bouwens et al.}{2003b}]{bou03b}
Bouwens, R. J. et al. 2003, \apj, 595, 589

\bibitem[\protect\citeauthoryear{Bower, Lucey \& Ellis}{1992}]{bow92} 
Bower, R. G., Lucey, J. R., \& Ellis, R. S. 1992, \mnras, 254, 601

\bibitem[\protect\citeauthoryear{Brinchmann et al.}{2004}]{bri04}
Brinchmann, J., Charlot, S., White, S. D. M., Tremonti, C., Kauffmann, G., Kechman, T., \& Brinkmann 2004, \mnras, 351, 1151

\bibitem[\protect\citeauthoryear{Brinchmann \& Ellis}{2000}]{bri00}
Brinchmann, J. \& Ellis, R. 2000, \apj, 536, L77

\bibitem[\protect\citeauthoryear{Bullock et al.}{2001}]{bul01}
Bullock, J. S., Dekel, A., Kolatt, T. S., Kravtsov, A. V., Klypin, A. A., Porciani, C., Primack, J. R. 2001, \apj, 555, 240

\bibitem[\protect\citeauthoryear{Bunker et al.}{2004}]{bun04}
Bunker, A, Stanway, E. R., Ellis, R. S., McMahon, R. G. 2004, \mnras, 355, 374

\bibitem[\protect\citeauthoryear{Carraro, Lia \& Chiosi}{1998}]{car98}
Carraro, G. Lia, C., \& Chiosi, C. 1998, \mnras, 297, 1021

\bibitem[\protect\citeauthoryear{Cen \& Ostriker}{1999}]{cen99} 
Cen, R. \& Ostriker, J. P. 1999, \apj, 519, L109

\bibitem[\protect\citeauthoryear{Cohen}{2002}]{coh02} 
Cohen, J. G. 2002, \apj, 567, 672

\bibitem[\protect\citeauthoryear{Cole et al.}{1994}]{col94} 
Cole, S., Arag\'on-Salamanca, A., Frenk, C. S., Navarro, J. F., \& Zepf, S. E. 
1994, \mnras, 271, 781

\bibitem[\protect\citeauthoryear{Cole et al.}{2001}]{col01} 
Cole, S. et al. 2002, \mnras, 326, 255

\bibitem[\protect\citeauthoryear{Connolly et al.}{1997}]{con97} 
Connolly, A. J., Szalay, A. S., Dickinson, M., SubbaRao, M. U., \&
 Brunner, R. J. 1997, \apj, 486, L11

\bibitem[\protect\citeauthoryear{Cowie et al.}{1996}]{cow96}
Cowie, L. L., Songaila, A., Hu, E. M., \& Cohen, J. G. 1996, \aj, 112, 839

\bibitem[\protect\citeauthoryear{Croton et al.}{2006}]{cro06}
Croton, D. et al. 2006, \mnras, 365, 11

\bibitem[\protect\citeauthoryear{Dav\'{e} \& Oppenheimer}{2006}]{dav06}
Dav\'{e}, R. \& Oppenheimer, B. D. 2006, astro-ph/0608268

\bibitem[\protect\citeauthoryear{Dickinson et al.}{2003}]{dic03}
Dickinson, M., Papovich, C., Ferguson, H. C., \& Budav\'{a}ri, T. 2003, \apj, 587, 25

\bibitem[\protect\citeauthoryear{Di Matteo, Springel, \& Hernquist}{2005}]{dim05} 
Di Matteo, T., Springel, V., \& Hernquist, L. 2005, Nature, 433, 604

\bibitem[\protect\citeauthoryear{Erb et al.}{2006}]{erb06}
Erb, D. K., Shapley, A. E., Pettini, M., Steidel, C. C., Reddy, N. A. \& Adelberger, K. L. 2006, \apj, 644, 813

\bibitem[\protect\citeauthoryear{Fontana et al.}{2006}]{fon06}
Fontana, A. et al. \aap, 459, 745

\bibitem[\protect\citeauthoryear{F{\"o}rster Schreiber et al.}{2004}]{fos04}
F{\"o}rster Schreiber, N. M. et al. 2004, \apj, 616, 40

\bibitem[\protect\citeauthoryear{Fukugita \& Peebles}{2004}]{fuk04}
Fukugita, M. \& Peebles, P. J. E. 2004, \apj, 616, 643

\bibitem[\protect\citeauthoryear{Gallego et al.}{1995}]{gal95}
Gallego, J., Zamorand, J., Arag\'on-Salamanca, A., \& Rego, M. 1995, 
\apj, 455, L1

\bibitem[\protect\citeauthoryear{Gallazzi et al.}{2005}]{gal05}
Gallazzi, A., Charlot, S., Brinchmann, J., White, S. D. M., \& Tremonti, C. A. 2005, \mnras, 362, 41

\bibitem[\protect\citeauthoryear{Giavalisco et al.}{2004}]{gia04}
Giavalisco, M et al. 2004, \apj, 600, L103

\bibitem[\protect\citeauthoryear{Glazebrook et al.}{1999}]{gla99}
Glazebrook, K., Blake, C., Economou, F., Lilliy S., \& Colles, M.  1999,
\mnras, 306, 843

\bibitem[\protect\citeauthoryear{Gronwall}{1999}]{gro99}
Gronwall, C. 1999, in AIP Conf. Proc. 470, After the Dark Ages: When Galaxies Were Young, ed. S. Holt \& E. Simith (New York: AIP), 335

\bibitem[\protect\citeauthoryear{Haardt \& Madau}{1996}]{haa96}
Haardt, F. \& Madau, P. 1996, \apj, 461, 20

\bibitem[\protect\citeauthoryear{Heckman et al.}{2000}]{hec00}
Heckman, T. M., Lehnert, M. D., Strickland, D. K.,\&  Armus, L. 2000, \apjs, 129, 492

\bibitem[\protect\citeauthoryear{Mihos \& Hernquist}{1989}]{mih94}
Mihos, J. C. \&  Hernquist, L. 1994, \apj, 464, 641

\bibitem[\protect\citeauthoryear{Hughes et al.}{1998}]{hug98}
Hughes, D., et al. 1998, Nature, 394, 241

\bibitem[\protect\citeauthoryear{Iwata et al.}{2003}]{iwa03}
Iwata, I et al. 2003, \pasj, 55, 415

\bibitem[\protect\citeauthoryear{Kawata \& Gibson}{2003}]{kaw03}
Kawata, D., \& Gibson, B. K. 2003, \mnras, 340, 908

\bibitem[\protect\citeauthoryear{Katz}{1992}]{kat92}
Katz, N. 1992, \apj, 391, 502

\bibitem[\protect\citeauthoryear{Katz, Weinberg \& Hernquist}{1996}]{kat96}
Katz, N., Weinberg, D. H., \& Hernquist, L. 1996, \apjs, 105, 19 

\bibitem[\protect\citeauthoryear{Kauffmann, White \& Guiderdoni}{1993}]{kau93} 
Kauffmann, G., White, S. D. M., \& Guiderdoni, B. 1993, \mnras, 264, 201

\bibitem[\protect\citeauthoryear{Kauffmann et al.}{2003}]{kau03} 
Kauffmann, G., et al. 2003, \mnras, 341, 54

\bibitem[\protect\citeauthoryear{Kobayashi}{2004}]{kob04} 
Kobayashi, C., 2004, \mnras, 347, 740 (K04)

\bibitem[\protect\citeauthoryear{Kobayashi}{2005}]{kob05} 
Kobayashi, C., 2005, \mnras, 361, 1216

\bibitem[\protect\citeauthoryear{Kobayashi \& Arimoto}{1999}]{kob99} 
Kobayashi, C., \& Arimoto, N. 1999, \apj, 527, 573

\bibitem[\protect\citeauthoryear{Kobayashi, Tsujimoto \& Nomoto}{2000}]{kob00} 
Kobayashi, C., Tsujimoto, T., \& Nomoto, K. 2000, \apj, 539, 26

\bibitem[\protect\citeauthoryear{Kobayashi et al.}{1998}]{kob98} 
Kobayashi, C., Tsujimoto, T., Nomoto, K., Hachisu, I, \& Kato, M. 1998, \apj, 503, L155

\bibitem[\protect\citeauthoryear{Kobayashi et al.}{2006}]{kobnom06} 
Kobayashi, C., Umeda, H., Nomoto, K., Tominaga, N., \& Ohkubo, T. 2006, \apj, 653, 1145

\bibitem[\protect\citeauthoryear{Kobulnicky \& Kewley}{2004}]{kobu04} 
Kobulnicky, H. A. \& Kewley, L. J. 2004, \apj, 617, 240

\bibitem[\protect\citeauthoryear{Kodama \& Arimoto}{1997}]{kod97} 
Kodama, T., \& Arimoto, N. 1997, \aap, 320, 41

\bibitem[\protect\citeauthoryear{Kodama et al.}{2004}]{kod04} 
Kodama, T., et al. 2004, \mnras, 350, 1005

\bibitem[\protect\citeauthoryear{Kulkarni et al.}{2005}]{kul05}
Kulkarni,  V. P., Fall, S. M., Lauroesch, J. T., York, D. G., Welty, D. E., Khare, P. \& Truran, J. W. 2005, \apj, 618, 68

\bibitem[\protect\citeauthoryear{Larson}{1974}]{lar74}
Larson, R. B. 1974, \mnras, 169, 229

\bibitem[\protect\citeauthoryear{Lilly et al.}{1996}]{lil96}
Lilly, S. J., Le F\`evre, O., Hammer, F., \& Crampton, D. 1995, \apj, 460, L1

\bibitem[\protect\citeauthoryear{McGaugh}{2005}]{mcg05}
McGaugh, S. S., 2005, \apj, 632, 859

\bibitem[\protect\citeauthoryear{Madau et al.}{1996}]{mad96} 
Madau, P., Ferguson, H. C., Dickinson, M. E., Giavalisco, M., Steidel, C. C., 
\& Fruchter, A. 1996, \mnras, 283, 1388

\bibitem[\protect\citeauthoryear{Martin, Kobulnicky \& Heckman}{2002}]{mar02}
Martin, C. L., Kobulnicky, H. A., \& Heckman, T. M. 2002 \apj, 574, 663 

\bibitem[\protect\citeauthoryear{Mazzotta et al.}{2002}]{maz02}
Mazzotta, P., Kaastra, J. S., Paerels, F. B., Ferrigno, C., Colafrancesco, S., Mewe, R., Forman, W. R. 2002, \apj, 567, 37 

\bibitem[\protect\citeauthoryear{Mosconi et al.}{2001}]{mos01}
Mosconi, M. B., Tissera, P. B., Lambas, D. G., \& Cora, S. A., 2001, \mnras, 325, 34

\bibitem[\protect\citeauthoryear{Nakasato \& Nomoto}{2003}]{nak03}
Nakasato, N., \& Nomoto, K. 2003, \apj, 588, 842

\bibitem[\protect\citeauthoryear{Navarro, Frenk \& White}{1996}]{nav96}
Navarro, J. F., Frenk, C. S., \& White, S. D. M. 1996, \apj, 462, 563

\bibitem[\protect\citeauthoryear{Navarro \& White}{1993}]{nav93}
Navarro, J. F., \& White, S. D. M. 1993, \mnras, 265, 271

\bibitem[\protect\citeauthoryear{Navarro \& White}{1994}]{nav94}
Navarro, J. F., \& White, S. D. M. 1994, \mnras, 267, 401

\bibitem[\protect\citeauthoryear{Nomoto et al.}{1997a}]{nom97a}
Nomoto, K., Hashimoto, M, Tsujimoto, T, Thielemann, F. -K, Kishimoto, N., Kubo, Y., \&  Nakasato, N. 1997a, Nuclear Physics, A616, 79c

\bibitem[\protect\citeauthoryear{Nomoto et al.}{1997b}]{nom97b}
Nomoto, K., Iwamoto, K., Nakasato, N., Thielemann, F. -K, Brachwitz, F., Tsujimoto, T., Kubo, Y., \&  Kishimoto, N. 1997b, Nuclear Physics, A621, 467c

\bibitem[\protect\citeauthoryear{Nomoto et al.}{2002}]{nom02}
Nomoto, K., Maeda, K., Umeda, H., Ohkubo, T., Deng, J., \& Mazzali, P.
2002, in IAU Symp. 212, in press (astro-ph/0209064)

\bibitem[\protect\citeauthoryear{Norman et al.}{2004}]{nor04}
Norman, C. et al. 2004, \apj, 607, 721

\bibitem[\protect\citeauthoryear{Ohyama et al.}{2002}]{ohy02}
Ohyama, Y, et al. 2002, PASJ, 54, 891

\bibitem[\protect\citeauthoryear{Ouchi et al.}{2004}]{ouc04}
Ouchi, M, et al. 2004, \apj, 611, 660

\bibitem[\protect\citeauthoryear{Pahre, Djorgovski \& de Carvalho}{1998}]{pah98}
Pahre, M., Djorgovski, S. G., \& de Carvalho, R. R. 1998, \apj, 116, 1591

\bibitem[\protect\citeauthoryear{P\'{e}rez-Gonz\'{a}lez et al.}{2003}]{per03} 
P\'{e}rez-Gonz\'{a}lez, P. G., Zamorano, J., Gallego, J., Arag\'{o}n-Salamanca, A., \& Gil de Paz, A. 2003, \apj, 591, 827

\bibitem[\protect\citeauthoryear{Pettini et al.}{1997}]{pet97} 
Pettini, M., Smith, L. J., King, D. L., \& Hunstead, R. W. 1997, \apj, 486, 665

\bibitem[\protect\citeauthoryear{Pettini et al.}{2001}]{pet01}
Pettini, M., et al. 2001, \apj, 554, 981

\bibitem[\protect\citeauthoryear{Pfenniger \& Revaz}{2005}]{pfe05}
Pfenniger, D. \& Revaz, Y. 2005, \aap, 431, 511

\bibitem[\protect\citeauthoryear{Prochaska et al.}{2003}]{pro03}
Prochaska, J., Gawiser, E., Wolfe, A. M., Castro, S., \& Djorgovski, S. G. 2003, \apj, 595, L9

\bibitem[\protect\citeauthoryear{Raiteri, Villata \& Navarro}{1996}]{rai96}
Raiteri, C. M., Villata, M., \& Navarro, J. F., \aap, 315, 105

\bibitem[\protect\citeauthoryear{Reddy \& Steidel}{2004}]{red04}
Reddy, N. A. \& Steidel, C. C. 2004, \apj, 603, L13

\bibitem[\protect\citeauthoryear{Renzini}{2002}]{ren02}
Renzini, A. 2002, in ASP Conference Series, 253, Chemical Enrichment of Intracluster and Intergalactic Medium, eds., R. Fusco-Femiano \& F. Matteucci, p.331

\bibitem[\protect\citeauthoryear{Rudnick et al.}{2003}]{rud03}
Rudnick, G. et al. 2003, \apj, 599, 847

\bibitem[\protect\citeauthoryear{Rudnick et al.}{2006}]{rud06}
Rudnick, G. et al. 2006, \apj, 650, 624

\bibitem[\protect\citeauthoryear{Schaye et al.}{2003}]{sch03} 
Schaye, J., Aguirre, A., Kim, T., Theuns, T., Rauch, M., \& Sargent, W. L. W. 2003, \apj, 596, 768

\bibitem[\protect\citeauthoryear{Schiminovich et al.}{2005}]{sch05}
Schiminovich, D. et al. 2005, \apj, 619, L47

\bibitem[\protect\citeauthoryear{Songaila}{2001}]{son01} 
Songaila, A. 2001, \apj, 561, L153

\bibitem[\protect\citeauthoryear{Springel}{2005}]{spr05} 
Springel, V. 2005, \mnras, 364, 1105

\bibitem[\protect\citeauthoryear{Springel \& Hernquist}{2002}]{spr02} 
Springel, V. \& Hernquist, L. 2002, \mnras, 333, 739 

\bibitem[\protect\citeauthoryear{Springel \& Hernquist}{2003}]{spr03} 
Springel, V. \& Hernquist, L. 2003, \mnras, 339, 289 

\bibitem[\protect\citeauthoryear{Springel \& Hernquist}{2003b}]{spr03b} 
Springel, V. \& Hernquist, L. 2003, \mnras, 339, 312 

\bibitem[\protect\citeauthoryear{Springel, Yoshida \& White}{2000}]{spr00} 
Springel, V., Yoshida, N.,  \& White, S. D. M. 2000, NewA, 6, 79

\bibitem[\protect\citeauthoryear{Steidel et al.}{1999}]{ste99}
Steidel, C. C., Adelberger, K. L., Giavalisco, M., Dickinson, M., \& Pettini, M. 1999, \apj, 519, 1

\bibitem[\protect\citeauthoryear{Steinmetz \& M\"uller}{1994}]{ste94}
Steinmetz, M., \& M\"uller, E. 1994, \aap, 281, L97

\bibitem[\protect\citeauthoryear{Sutherland \& Dopita}{1993}]{sut93}
Sutherland, R. S., \& Dopita, M. A. 1993, \apjs, 88, 235

\bibitem[\protect\citeauthoryear{Thomas \& Maraston}{2003}]{tho03}
Thomas, D. \& Maraston, C. 2003, \aap, 401, 429

\bibitem[\protect\citeauthoryear{Tornatore et al.}{2004}]{tor04} 
Tornatore, L., Borgani, S., Matteucci, F., Recchi, S., Tozzi, P. 2004, \mnras, 349, L19

\bibitem[\protect\citeauthoryear{Tresse \& Maddox}{1998}]{tres98}
Tresse, L., \& Maddox, S. J. 1998, \apj, 495, 691

\bibitem[\protect\citeauthoryear{Tresse et al.}{2002}]{tre02}
Tresse, L., Maddox, S. J., Le F\`{e}vre, \& Cuby, J.-G. 2002, \mnras, 337, 369

\bibitem[\protect\citeauthoryear{Tremonti et al.}{2004}]{tre04}
Tremonti, C. A. et al. 2004, \apj, 613, 898

\end{thebibliography}
\end{document}